\documentclass[runningheads]{llncs}
\usepackage[T1]{fontenc}
\usepackage{graphicx}
\usepackage{amsmath,graphicx}
\usepackage{algorithm, algorithmicx, algpseudocode}
\usepackage{tabularx}
\usepackage{multirow}
\usepackage{pifont}
\usepackage{caption}
\usepackage{subcaption}
\usepackage[export]{adjustbox}
\usepackage{graphics}
\usepackage{array}
\usepackage{xcolor}
\usepackage{amssymb}
\usepackage{booktabs}
\usepackage{color}
\usepackage{comment}
\usepackage{makecell}
\usepackage{adjustbox}
\usepackage{orcidlink}
	
\usepackage{cite}
\begin{document}
\title{Blind Super Resolution with Reference Images and Implicit Degradation Representation}
\titlerunning{RDSR}
%


\begingroup
\renewcommand{\thefootnote}{\fnsymbol{footnote}}
\setcounter{footnote}{1}
\footnotetext[1]{{Equal contributions}}
\endgroup

\begingroup
\renewcommand{\thefootnote}{\fnsymbol{footnote}}
\setcounter{footnote}{2}
\footnotetext[2]{{Corresponding Author}}
\endgroup

\author{Huu-Phu Do\inst{1}\protect\footnotemark[1]\orcidlink{0009-0006-7327-9016}\and
Po-Chih Hu\inst{1}\orcidlink{0009-0000-2384-284X}\protect\footnotemark[1] \and
Hao-Chien Hsueh\inst{1}\orcidlink{0009-0007-9925-9504} \and 
Che-Kai Liu\inst{1}\orcidlink{0009-0000-1101-8431} \and
Vu-Hoang Tran\inst{2}\orcidlink{0000-0002-1352-4853}  \and 
Ching-Chun Huang \inst{1}\orcidlink{0000-0002-4382-5083}\protect\footnotemark[2]
}

%
\authorrunning{Phu Do, Chih Hu\ et al.}
%
\institute{National Yang Ming Chiao Tung University, Taiwan \and
Ho Chi Minh City University of Technology and Education, Vietnam}

\maketitle              
\begin{abstract}
Previous studies in blind super-resolution (BSR) have primarily concentrated on estimating degradation kernels directly from low-resolution (LR) inputs to enhance super-resolution. However, these degradation kernels, which model the transition from a high-resolution (HR) image to its LR version, should account for not only the degradation process but also the downscaling factor. Applying the same degradation kernel across varying super-resolution scales may be impractical. Our research acknowledges degradation kernels and scaling factors as pivotal elements for the BSR task and introduces a novel strategy that utilizes HR images as references to establish scale-aware degradation kernels. By employing content-irrelevant HR reference images alongside the target LR image, our model adaptively discerns the degradation process. It is then applied to generate additional LR-HR pairs through down-sampling the HR reference images, which are keys to improving the SR performance. Our reference-based training procedure is applicable to proficiently trained blind SR models and zero-shot blind SR methods, consistently outperforming previous methods in both scenarios. This dual consideration of blur kernels and scaling factors, coupled with the use of a reference image, contributes to the effectiveness of our approach in blind super-resolution tasks.
\end{abstract}

\section{Introduction}
\label{sec:intro}

Image super-resolution (SR) is a widely studied task in the field of image processing, aiming to enhance the resolution and quality of low-resolution (LR) images. Extracting detailed information from an LR image for super-resolution poses significant challenges. Previous researchers have employed deep learning techniques to improve the quality of LR images \cite{DCNSR, deeply2016, embedded2019, EDSR}. This task typically requires a substantial amount of high-resolution (HR) images and their corresponding LR counterparts for training an SR network through a supervised approach. Prior studies often utilized a bicubic kernel to synthesize low-resolution images from high-resolution ones \cite{EDSR}. This process can be formulated as:
\begin{equation}
    {I}^{LR} = ({I}^{HR} \otimes k) \downarrow_{s} + n,
\end{equation}
where ${I}^{LR}$ denotes the low-resolution image, ${I}^{HR}$ is the high-resolution image, $k$ represents the degradation kernel, $\downarrow_{s}$ represents downsampling operation with scale factor s and $n$ is the added noise. 

\begin{figure}[tb]
  \centering
  \begin{subfigure}{0.45\linewidth}
    \includegraphics[height=2.7cm]{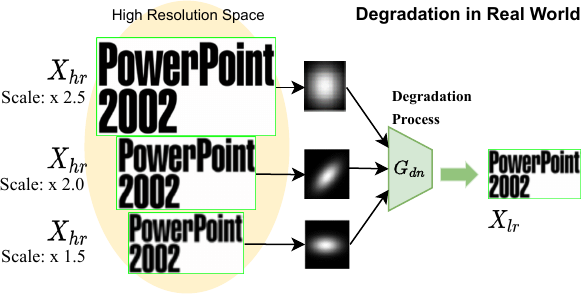}
    \caption{Degradation in Real World}
    \label{fig:motivation-a}
  \end{subfigure}
  \begin{subfigure}{0.45\linewidth}
    \centering
    \includegraphics[height=1.4cm]{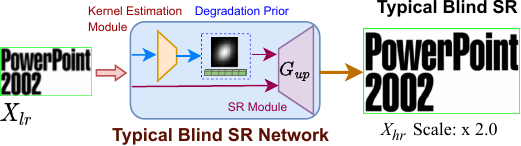}
    \caption{Typical Blind SR}
    \label{fig:motivation-b}
  \end{subfigure}

  \begin{subfigure}{0.9\linewidth}
    \centering
    \includegraphics[scale=0.9]{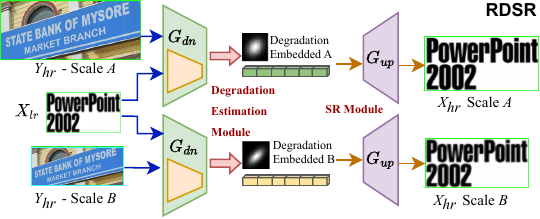}
    \caption{RDSR}
    \label{fig:motivation-c}
  \end{subfigure}
  \caption{(a) Degradation with multi scales in real world (b) Typical Blind Super-Resolution (c) Proposed Method - RDSR.}
  \label{fig:motivation-all}
\end{figure}

In practical applications, blur kernels are often much more complex than the typically bicubic ones. This domain gap can result in significant performance degradation when applying these networks to real-world scenarios. Therefore, there is a critical need to focus on super-resolution (SR) techniques that can handle unknown blur kernels, known as blind SR. Currently, existing blind SR techniques \cite{IKC, KernelGAN} primarily focus on deriving a single degradation kernel that is assumed to be universally applicable across different SR scales. As depicted in Fig. \ref{fig:motivation-all}(b), predominantly involve estimating kernels for a specific scaling factor and incorporating them within the super-resolution (SR) model. Consequently, thereby overlook the nuances introduced by varying scales. 
Effective blind super-resolution, however, requires accounting for the variability in blur kernels and the ability to adapt to different scaling factors.
As illustrated in Fig. \ref{fig:motivation-all}(a), the degradation kernel associated with a specific scaling factor significantly differs from kernels tailored for other scaling factors, emphasizing the need for a more flexible and scale-aware approach for kernel estimation.

Therefore, in this paper, we propose an approach for a degradation kernel that is inherently scale-aware. This is crucial for accurately super-resolving a low-resolution (LR) image to its high-resolution (HR) counterparts across various scales, necessitating a degradation kernel that is precisely tailored for each specific scenario. 
To achieve this goal, we introduce a reference-based training procedure that integrates content-irrelevant HR reference images into the super-resolution (SR) model as conditional information. This innovative approach allows the SR model to leverage diverse high-resolution data, enhancing its ability to reconstruct high-quality images from low-resolution inputs. These reference images serve dual purposes: (a) to introduce a scaling factor as a condition for the SR network, and (b) to provide additional HR information, thereby facilitating image enhancement at the desired resolution. Fig. \ref{fig:motivation-all}(c) depicts the overall concept of our method.

Diverging from existing reference-based SR approaches \cite{C2Matching,MASA,TTSR} , our methodology does not directly utilize detailed information from the reference content; instead, it adaptively estimates a scale-aware kernel to model the degradation process accurately from the resolution of the reference images to that of the LR input. Our proposed procedure, termed ``\textbf{Reference Degradation Super Resolution (RDSR)}'' consists of upsampling and downsampling networks and is adaptable to various blind SR networks and zero-shot super-resolution networks. We also evaluate our instance implementation, the RDSR model, which utilizes KernelGAN\cite{KernelGAN} and a well-trained DASR\cite{DASR} as the backbones for the upsampling and downsampling networks respectively, and achieve the superior performance compared to traditional SR methods. The main findings and contributions of this paper are summarized as follows:
\begin{enumerate}
\item The degradation processes for an LR image derived from HR images with different resolutions should be distinct. This suggests that the corresponding blur kernel from HR images downsampled to varying resolutions should also differ.
\item Our proposed approach integrates reference images into the architecture to facilitate scaling information for refining the degradation kernel, thereby optimizing the performance of the super-resolution network.
\item In contrast to other reference-based super-resolution methods that rely on learning high-resolution (HR) content from closely related reference images, our approach permits the use of content-irrelevant HR reference images. This strategy effectively mitigates the risk of content inconsistency issues, which are prevalent in traditional reference-based SR methods.
\item We introduce adversarial and regularization losses to optimize our proposed network. The adversarial loss encourages the upsampling network to align with the domain distribution of the reference images. Meanwhile, the regularization loss ensures consistency in degradation priors between the reference and target images.
\item  Our reference-based training procedure is specifically designed for the target LR image and operates independently of other LR images. This approach is applicable not only to well-trained blind SR models but also to zero-shot super-resolution models.
\end{enumerate}

\section{Related Work}\label{sec:related}
Numerous studies \cite{deepunfolding, MDSR} focus on single image super-resolution (SISR) tasks with known degradation. However, these methods may experience significant performance drops when the degradation process is misaligned with the training samples. In addressing blind SR, various approaches have been proposed, categorized into four main groups: degradation estimation methods, unsupervised methods, zero-shot methods, and reference-based methods.

\subsection{Degradation Estimation of Blind SR}
To tackle image super-resolution with unknown degradation, numerous studies concentrate on degradation estimation. The objective is to uncover clues within the LR images and utilize the estimated degradation to guide the super-resolution process. These methods fall into two categories: explicit kernel estimation and implicit kernel estimation.

\subsubsection{Explicit Kernel Estimation}
The explicit kernel estimation methods involve directly evaluating a degradation kernel, as demonstrated in previous works such as IKC \cite{IKC} and KernelGAN \cite{KernelGAN}. IKC introduced an iterative method for correcting the kernel based on intermediate SR images during the training stage. However, this correction method requires multiple degradation kernels, along with corresponding LR and HR images during training. KernelGAN proposed a kernel estimation approach that learns the degradation kernel by establishing internal connections between patches in a single LR image. This sets it apart from other SR methods that rely on external training materials. This attribute not only offers flexibility but is also precisely customized for the target images.

\subsubsection{Implicit Kernel Estimation}
Different from explicit kernel estimation which may suffer a performance drop if the kernel estimation error increases, DASR \cite{DASR} and CDSR \cite{CDSR} proposed implicit kernel estimation for learning the degradation representations using contrastive learning. 
The kernel estimator produces a degradation representation, serving as an implicit degradation prior, and this representation is seamlessly integrated into the SR network to guide the model through various degradation scenarios.

\subsection{Unsupervised Super-Resolution Methods}
Unsupervised SR methods \cite{UPSSR, USVSR} distinguish themselves from supervised approaches that rely on paired LR and HR data with synthetic degradation kernels. Instead, these methods focused on learning the distribution within the domain encompassing bicubic and real-world degradation processes. Following the acquisition of domain distribution knowledge, a domain corrector is employed to enhance SR performance. 

\subsection{Zero-shot Super-Resolution Methods}
While the majority of SR methods necessitate paired data comprising LR, HR, and degradation kernel information for training, this approach poses challenges in terms of adaptability for individual images. In contrast, zero-shot super-resolution methods, exemplified by \cite{ZSSR, DualSR, RZSR} train models using only a single target LR image. These approaches yield an SR model tailored to the specific characteristics of the target images, deviating from the conventional generalized SR models. In the context of this, DualSR proposes a dual-path architecture optimized for blind super-resolution using only the input LR image. RZSR, on the other hand, advocates the creation of a reference dataset through cross-scale matching using the input target image. Our RDSR also targeted at this task. 

\subsection{Reference-based Super-Resolution Methods}
Reference-based super-resolution is a subfield within the field of super-resolution. In the prior studies, exemplified by \cite{TTSR, MASA, C2Matching}, the input comprises a target LR image along with a corresponding reference HR image. Notably, the content of the reference HR image is typically closely associated with the content of the target LR image. To leverage this relationship, most methods in this category utilize a feature-matching procedure to identify or evaluate related patches between the reference and target images. Once corresponding patches are identified, the SR procedure learns high-resolution information directly from the matched patches in the reference image. These methods assume a strong correlation between the reference and target images. However, inaccuracies in the matching process may lead to unexpected results. Additionally, the creation of paired target and reference images imposes additional costs.

\section{Proposed Method}
\label{sec:method}

We seek to harness the advantages of employing reference HR images in addressing zero-shot blind SR tasks. The following section provides an overview of our design and showcases instances of implementation. It illustrates the process of integrating reference images into our network based on the target image.
Furthermore, our training process only uses the target LR image and the provided irrelevant reference HR images. Specifically, our method yields an SR model tailored to the target image, making it a zero-shot super-resolution method.

\begin{figure}[!ht]
  \centering
  \includegraphics[width=0.9\linewidth]{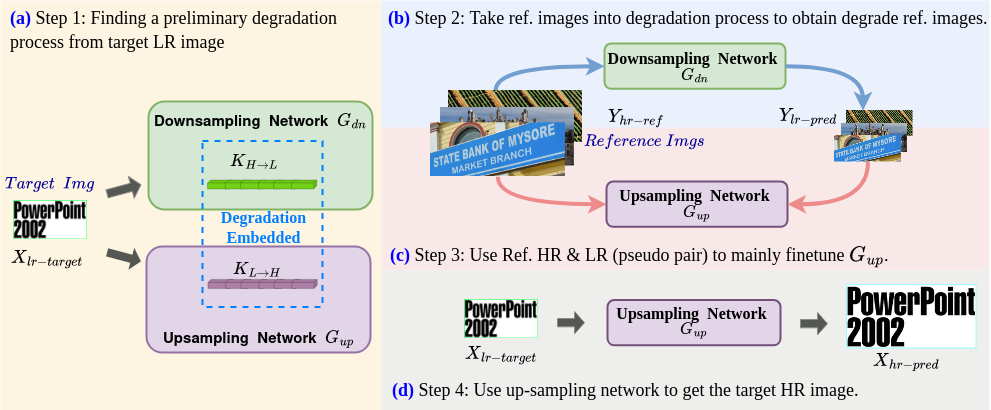}
  \caption{Overview of our Reference Degradation Super-Resolution Architecture. The diagram illustrates the RDSR framework, showcasing the sequential steps. (a) The degradation estimator first identifies the preliminary degradation process. (b) Reference images are introduced to capture the scale-aware degradation process. (c)(d)The reference image pairs, generated by the downsampling network, serve as a supervised signal to train the SR network, culminating in the production of high-quality super-resolution results.
  }
  \label{fig:rdsr-concept}
\end{figure}


\subsection{Overview and Pipeline} \label{Overview and Pipeline}

We propose a novel procedure to optimize the super-resolution task. As illustrated in Fig. \ref{fig:rdsr-concept}, our method involves four key steps. The first step (Fig. \ref{fig:rdsr-concept}(a)) involves  identifying a preliminary degradation process from the target LR image $X_{lr-tar}$. This degradation process is embedded and can be represented either implicitly or explicitly in downsampling network $G_{dn}$ and upsampling network $G_{up}$, denoted by $K_{H{\rightarrow}L}$ and $K_{L{\rightarrow}H}$, respectively. Specifically, $K_{H{\rightarrow}L}$ performs the mapping from HR to LR image, while $K_{L{\rightarrow}H}$ performs the mapping from LR to HR image. 

Next, given the HR reference images, we aim to generate pseudo data through the downsampling network $G_{dn}$, which is initialized from the previous step, as depicted in Fig. \ref{fig:rdsr-concept}(b). Unlike existing methods that utilize HR images with content similar to the LR images, our approach leverages irrelevant HR images. These irrelevant HR images $Y_{hr-ref}$ are downscaled to generate corresponding LR images $Y_{lr-pred}$, which are then employed as a data augmentation technique during the training phase.

Subsequently, as illustrated in Fig. \ref{fig:rdsr-concept}(c), these pseudo paired data $Y_{hr-ref}$ and $Y_{lr-pred}$ are primarily employed for fine-tuning the upsampling network $G_{up}$. This process allows $G_{up}$ to progressively adapt to the reference image scale by refining the degradation representation $K_{L{\rightarrow}H}$ through constraints derived from the degradation of the target LR image and the downscaled reference HR images. This fine-tuning results in substantial improvements in super-resolution (SR) performance. Additionally, since the downsampling network $G_{dn}$ is specifically initialized for the LR target images, it is also simultaneously finely adjusted to align the degradation representation $K_{H{\rightarrow}L}$ with that of the upsampling network $K_{L{\rightarrow}H}$.

Finally, following comprehensive fine-tuning, the upsampling network $G_{up}$  is utilized for downstream tasks that take LR target image ($X_{lr-tar}$) as input and generate corresponding HR image ($X_{hr-pred}$), as illustrated in Fig. \ref{fig:rdsr-concept}(d). Importantly, our approach offers the flexibility to interchange the downsampling network $G_{dn}$ and the upsampling network $G_{up}$.

\subsection{Implementation}
\begin{figure}[!h]
  \centering
  \includegraphics[width=0.9\linewidth]{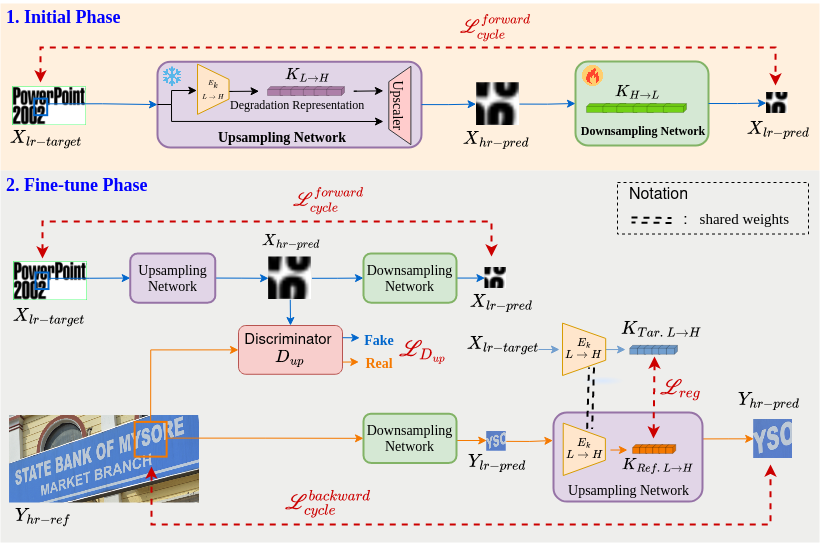}
  \caption{Overview of RDSR Implementation.
  }
  \label{fig:RDSR-flow}
\end{figure}

In this section, we describre an instance of our procedure in term RDSR model,  which utilizes KernelGAN \cite{KernelGAN} and DASR \cite{DASR} architecture as the backbone for downsampling network $G_{dn}$ and upsampling network $G_{up}$ respectively. Based on our procedure described at section \ref{Overview and Pipeline}, to simply, we divide the implementation into two parts: Initial Phase and Fine-tune Phase, as illustrated in Fig. \ref{fig:RDSR-flow}. 

In the initial phase as depicted in Fig. \ref{fig:RDSR-flow}(a), corresponding to Step 1, we aim to initialize the weights of $G_{up}$ and $G_{dn}$ to tailor them for the degradation process of the target LR image. Specifically, we employ a pre-trained DASR model \cite{DASR} $G_{up}$, which includes a kernel estimator $E_{k\ L{\rightarrow}H}$ to estimate the implicit degradation representation $K_{L{\rightarrow}H}$ and an upscaler network to generate the upscaled image based on the LR input and the obtained $K_{L{\rightarrow}H}$ . In contrast, KernelGAN \cite{KernelGAN} $G_{dn}$, which explicitly extracts the kernel to compute the degradation representation $K_{H{\rightarrow}L}$ for downsampling the image, is randomly initialized and roughly optimized by using target LR and target HR pairs. These paired data are generated by utilizing the pre-trained $G_{up}$ and the target LR image $X_{lr-tar}$ to produce the target HR image $X_{hr-pred}$.

In the finetune phase, corresponding to Steps 2, 3, and 4, our objective is to generate additional LR-HR pairs by downsampling content-irrelevant reference HR images and utilizing them as references to establish scale-aware degradation kernels. As illustrated in Figure \ref{fig:RDSR-flow}(b), $G_{up}$ and $G_{dn}$ form two branches: the target branch and the reference branch, creating a dual-path pipeline inspired by CycleGAN\cite{cycleGAN}, with shared weights.


In the reference branch, the HR input is initially downsampled by $G_{dn}$ and then upsampled back to the original scale by $G_{up}$. In contrast, the target branch first upsamples the target LR image using $G_{up}$ and then downscales it back to the original LR size using $G_{dn}$. This process is designed to integrate HR image details into the model through the reference branch and ensure that the model maintains optimal parameters for processing the target LR images through the target branch. These objectives can be addressed by incorporating consistency loss along with constraints on the degradation representation, helping the model establish scale-aware degradation based on the HR images.
Additionally, the discriminator $D_{up}$ is incorporated to  distinguish between real HR images and upscaled target LR images produced by $G_{up}$. Ideally, the $G_{up}$ and $G_{dn}$ should act as inverses of each other in the target and reference branches. Therefore, leveraging reference HR images at a specific scale as additional data enhances the super-resolution performance for downstream tasks on target LR images.

\subsection{Training Objective}
In this section, we outline our training objectives, which encompass implementing cycle consistency training and integrating both the target and reference branches. Additionally, we utilize the adversarial network $D_{up}$ to align the output of the upsampling network with the HR distribution. To maintain consistency in the degradation process across both branches, we introduce a regularization loss for the degradation representation. Detailed descriptions of these specific training objectives will be provided in the following paragraphs. Note that both LR and HR images are processed as patches throughout the entire process.
\subsubsection{Cycle consistency loss.}  The cycle  consistency loss $\mathcal{L}_{cycle}$  consists of forward and backward cycle consistency losses during the training process of RDSR. It comprises the Charbonnier loss \cite{LPSR} and the perceptual loss \cite{perceptual2016}, utilizing an ImageNet-pretrained VGG-19 network denoted as $\phi$(.).
The forward cycle is performed on the target branch to facilitate minimizing the disparity between the original target LR image x and the cycle-generated target LR image. Ideally, this implies that $G_{dn}(G_{up}(x))=x$, ensuring that the model maintains optimal parameters for the target images while injecting HR information. The forward cycle consistency loss can be formulated as follows:


\begin{equation}
\begin{split}
\label{loss_forward}
    \mathcal{L}_{cycle}^{forward} = \mathbb{E}_{x}[\sqrt{({G}_{dn}({G}_{up}(x))-x)^{2} + \epsilon^{2}} \\
    + ||\phi_{n}({G}_{dn}({G}_{up}(x))) - \phi_{n}(x)||_1],
\end{split}
\end{equation}





To establish scale-awareness effectively from HR images y into the model, applying the backward cycle consistency loss is crucial. Ideally this aims for $G_{up}(G_{dn}(y))=y$. This loss is performed on the reference branch and can be written as:
\begin{equation}
\begin{split}
    \mathcal{L}_{cycle}^{backward} = \mathbb{E}_{y}[\sqrt{({G}_{up}({G}_{dn}(y))-y)^{2} + \epsilon^{2}} \\
    + ||\phi_{n}({G}_{up}({G}_{dn}(y))) - \phi_{n}(y)||_1],
\end{split}
\end{equation}
where $\epsilon$ is set empirically to $10^{-6}$ during the experiments and $\phi_{n}(.)$ is the feature at layer n of loss network $\phi$. The total cycle consistency loss is then the sum of the forward and backward cycle consistency losses :
\begin{equation}
    \mathcal{L}_{cycle} = \lambda_{cycle}^{forward}\mathcal{L}_{cycle}^{forward} +\lambda_{cycle}^{backward}\mathcal{L}_{cycle}^{backward}.
\end{equation}

\subsubsection{Generative and Discriminator loss.} Taking inspiration from SRGAN \cite{SRGAN}, we design an adversarial loss for the upsampling network $G_{up}$ to encourage it to generate more detailed images similar to the reference image. The adversarial loss for the generator is defined as follows:
\begin{equation}
    \mathcal{L}_{gan} =  \min_{G_{up}}\mathbb{E}_{x}[({D}_{up}({G}_{up}(x))-1)^{2}].
\end{equation}

On the other hand, the discriminator $D_{up}$ aims to distinguish fake generated high-resolution from real reference high-resolution images. Its objective is defined as follows: 
\begin{equation}
    \mathcal{L}_{D_{up}} = \min_{D_{up}}(\mathbb{E}_{y}[({D}_{up}(y)-1)^{2}]  
    + \mathbb{E}_{x}[{D}_{up}({G}_{up}(x))^{2}]).
\end{equation}

\subsubsection{Regularization loss.} To ensure consistency in the degradation process between target image x and reference image y, we assume that the degradation representation $K_{L{\rightarrow}H}$ of the target LR image from the target branch should ideally be similar to that of the downscaled reference image in the reference branch, implying $E_{k\ L{\rightarrow}H}(x) = E_{k\ L{\rightarrow}H}({G}_{dn}(y))$. Therefore, we introduce a regularization loss $\mathcal{L}_{reg}$, computed with L1 distance in our framework.
\begin{equation}
    \mathcal{L}_{reg} = \mathbb{E}_{x,y}[||E_{k\ L{\rightarrow}H}(x)-E_{k\ L{\rightarrow}H}({G}_{dn}(y))||_1].
\end{equation}

Finally, the total loss is defined as the sum of all components:
\begin{equation}
    \mathcal{L}_{total} =  \mathcal{L}_{cycle} + \lambda_{reg} \mathcal{L}_{reg} + \mathcal{L}_{gan},
\end{equation}


where $\lambda$ values control the importance of the different loss
function terms.

\section{Experiments and Results}
\label{sec:exp}

\newcommand{\xmark}{\ding{53}}%

\subsection{Datasets and Implementation Details}
For evaluation of SR performance, we use the datasets DIV2KRK \cite{KernelGAN} as target images and DIV2K \cite{DIV2K} as reference images for our experiments. DIV2KRK contains 100 images, LR images blurred with randomly generated kernels, which are 11x11 anisotropic Gaussians. DIV2K contains 800 images and serves as a reference collection in our experiments.

Although our method implies recovering to multiple scaling factors, for better comparison with other methods, we adopt scales commonly used $\times$2 and $\times$4 to evaluate the results. To evaluate our method’s performance, we adopted PSNR (Peak Signal-to-Noise Ratio) on Y channel of the images after being converted to the YCbCr color space and SSIM (Structural Similarity Index Measure).

In our experiments, we train 3000 iterations with a frozen $G_{up}$ in the first stage to obtain the preliminary downsampling network $G_{dn}$. Subsequently, $G_{up}$ is unfrozen to train 500 iterations per reference image on $E_{k\ L{\rightarrow}H}$ and the upscaling network, respectively. Concurrently, $G_{dn}$  is fine-tuned with minor adjustments. The initial learning rates are set $2\times10^{-3}$ and $1\times10^{-5}$ for $G_{dn}$ and $G_{up}$ respectively. We utilize the Adam\cite{adam} optimizer with $\beta_{1}=0.25$ and $\beta_{2}=0.99$. Additionally, $\lambda_{cycle}^{target}$ is set to 5, $\lambda_{cycle}^{ref}$ to 1, and $\lambda_{reg}$ to 20 based on our experiments for the whole training. The process for a LR image requires approximately 30min on a single RTX 2080Ti GPU.

Our RDSR method can generate multiple results for a target image utilizing various input reference images. To improve the quality of our training, we monitor specific criteria throughout the training process. Specifically, we observe the cycle consistency loss of the target image, deciding when to preserve the network parameters. Furthermore, we integrate a no-reference quality assessment index BRISQUE\cite{brisque} to guide the decision of whether to save the output of the network.

\subsection{Experimental Results and Comparison}
Our model utilizes well-trained DASR \cite{DASR} models as the backbone. These DASR models consist of $\times$2 and $\times$4 settings, which are trained using isotropic and anisotropic kernels, respectively. 
As shown in Table \ref{tab:exp-BSR-based}, we have achieved superior results in terms of PSNR and SSIM compared to the baseline DASR in both isotropic and anisotropic kernel types at scales 2 and 4. 
Furthermore, our method surpasses Bicubic and typical zero-shot super-resolution methods such as DualSR \cite{DualSR}, RZSR\cite{RZSR} in PSNR.
Additionally, as demonstrated in Fig. \ref{fig:vision-BSR-based1}, while bicubic interpolation results in blurry and oversmoothed images, DASR \cite{DASR} yields more detailed results. Our method produces the sharpest and clearest results among them.  It excels in restoring fine details and achieves visually superior results compared to both the baseline DASR \cite{DASR} and zero-shot methods, owing to its effective scale awareness. 

\newcolumntype{M}[1]{>{\centering\arraybackslash}m{#1}}
\begin{table}[!htbp]
    \caption{Quantitative comparison on DIV2KRK, we evaluate both blind super-resolution methods and zero-shot super-resolution methods.}
    \label{tab:exp-BSR-based}
    \centering
    \scalebox{0.85}{
    \begin{tabular}{M{0.1\linewidth} | M{0.3\linewidth} M{0.15\linewidth} M{0.15\linewidth} } \toprule
    Scale & Method & PSNR-Y & SSIM \\ \midrule
    \multirow{7}*{$\times$2} & 
    Bicubic & 28.829 & 0.8139 \\
    ~ & DualSR \cite{DualSR} & 30.92 & \textbf{0.8728} \\
    ~ & RZSR \cite{RZSR} & 30.55 & 0.8704 \\ 
    \cmidrule(l){2-2}
    ~ & DASR \cite{DASR} $isotropic$ & 30.743 & 0.8586 \\
    ~ & \quad + RDSR (Ours) & \textbf{30.973} & \textbf{0.8638} \\
    \cmidrule(l){2-2}
    ~ & DASR \cite{DASR} $anisotropic$ & 31.051 & 0.8668 \\
    ~ & \quad + RDSR (Ours) & \textbf{31.161} & \textbf{0.8683} \\
    \midrule
    \midrule
    \multirow{5}*{$\times$4} & 
    Bicubic & 25.471 & 0.7088  \\ 
    \cmidrule(l){2-2}
    ~ & DASR \cite{DASR} $isotropic$ & 27.441 & 0.7635  \\
    ~ & \quad + RDSR (Ours) & \textbf{27.560} & \textbf{0.7663}  \\
    \cmidrule(l){2-2}
    ~ & DASR \cite{DASR} $anisotropic$ & 28.142 & 0.7769  \\
    ~ & \quad + RDSR (Ours) & \textbf{28.233} & \textbf{0.7794}  \\\bottomrule
    \end{tabular}
    }
\end{table}

\newcommand\imgsize{1.0\textwidth}
\newcommand\rowspace{-0.1em}

\newcommand\cellsize{0.165\textwidth}
\newcommand\cellasize{0.16\textwidth}
\begin{figure}[tb]
\begin{center}
\begingroup
\setlength{\tabcolsep}{0.2pt} 
\renewcommand{\arraystretch}{1} 
\centering
\scalebox{0.85}{
\begin{tabular}{ m{\cellsize} m{\cellsize} m{\cellsize} m{\cellsize} m{\cellsize} m{\cellsize}}

LR \centering & Bicubic \centering & DASR \centering & RDSR (Ours) & Ground Truth & \hspace{1.7em} Ref. \\

\begin{subfigure}{\cellasize}
\includegraphics[width=\imgsize]{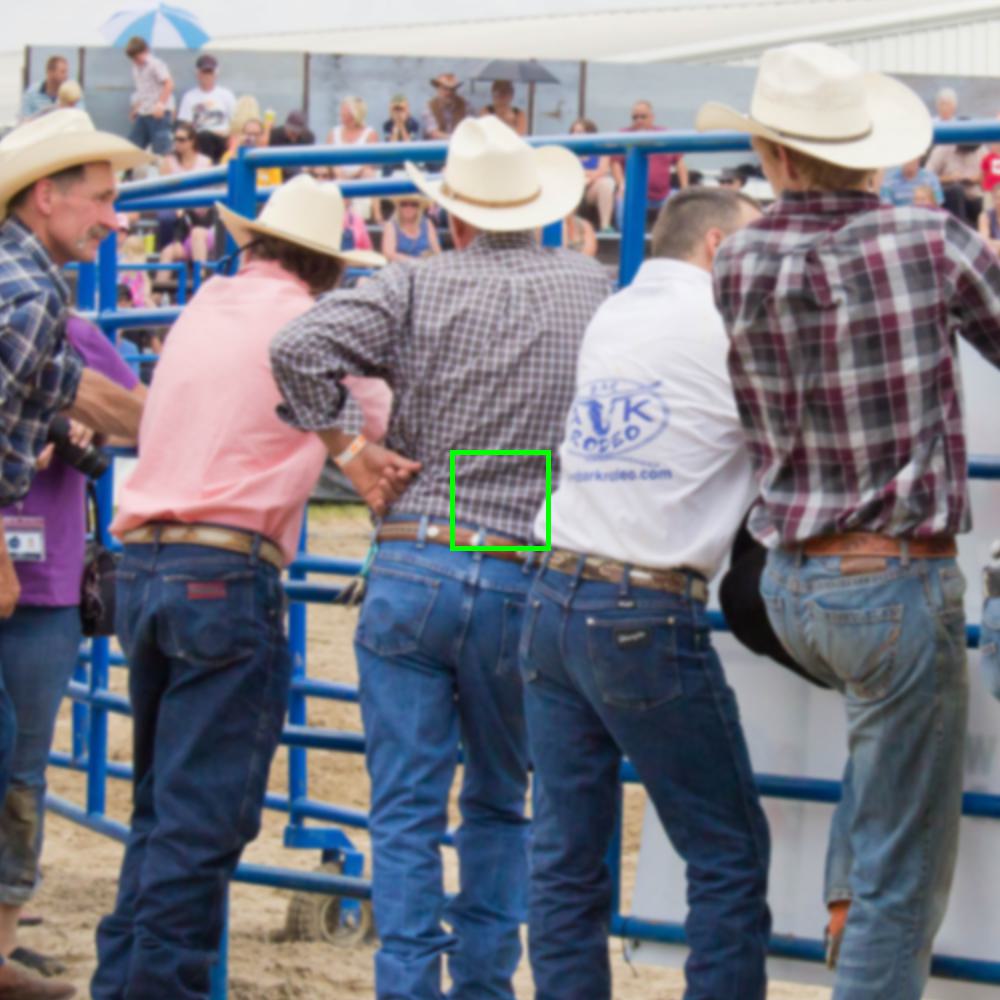}
\end{subfigure}
&
\begin{subfigure}{\cellasize}
\includegraphics[width=\imgsize]{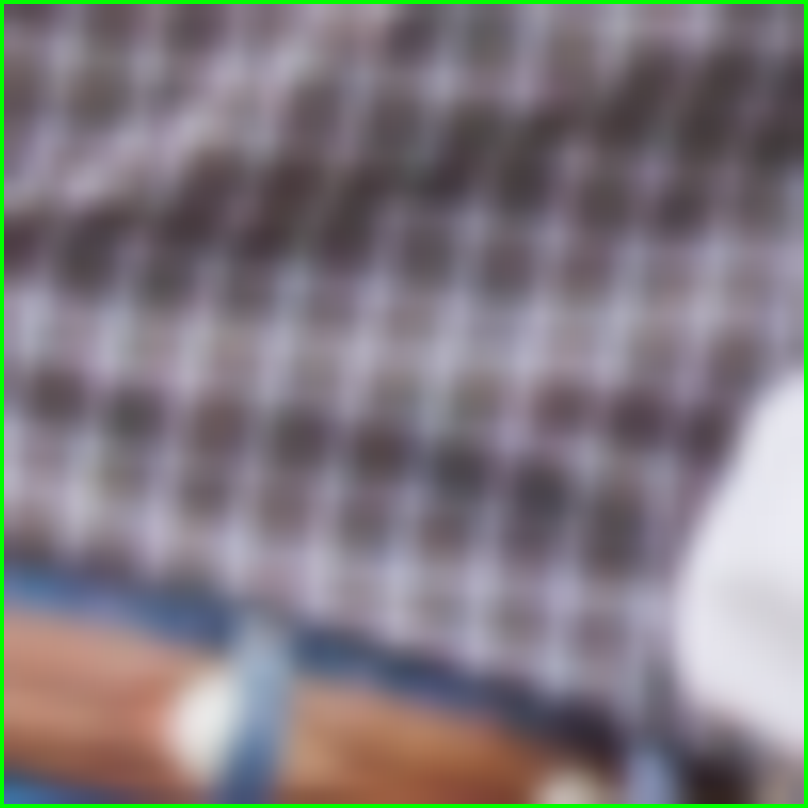}
\end{subfigure} 
&
\begin{subfigure}{\cellasize}
\includegraphics[width=\imgsize]{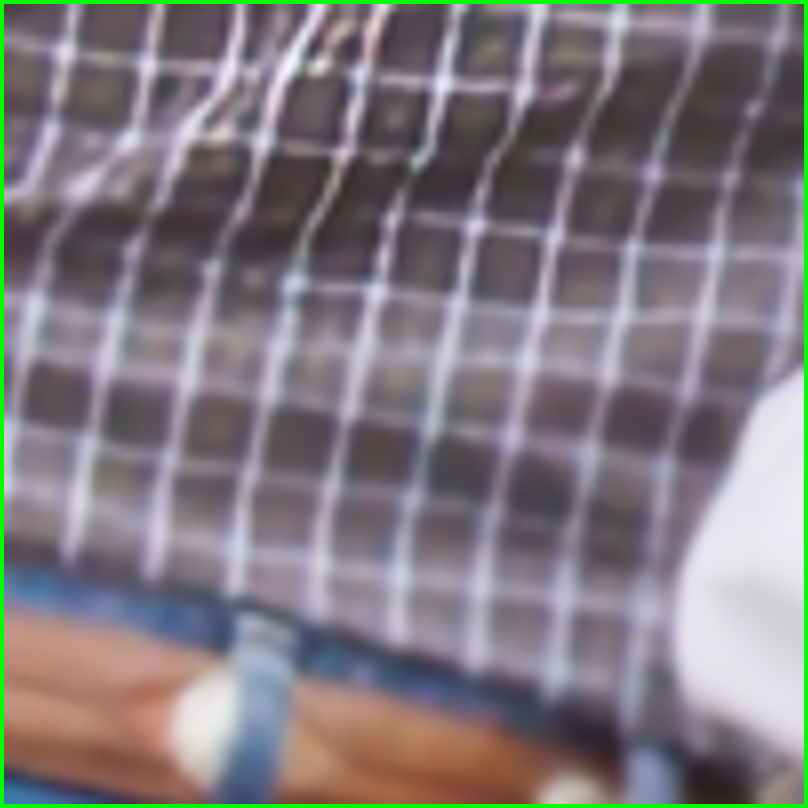}
\end{subfigure} 
&
\begin{subfigure}{\cellasize}
\includegraphics[width=\imgsize]{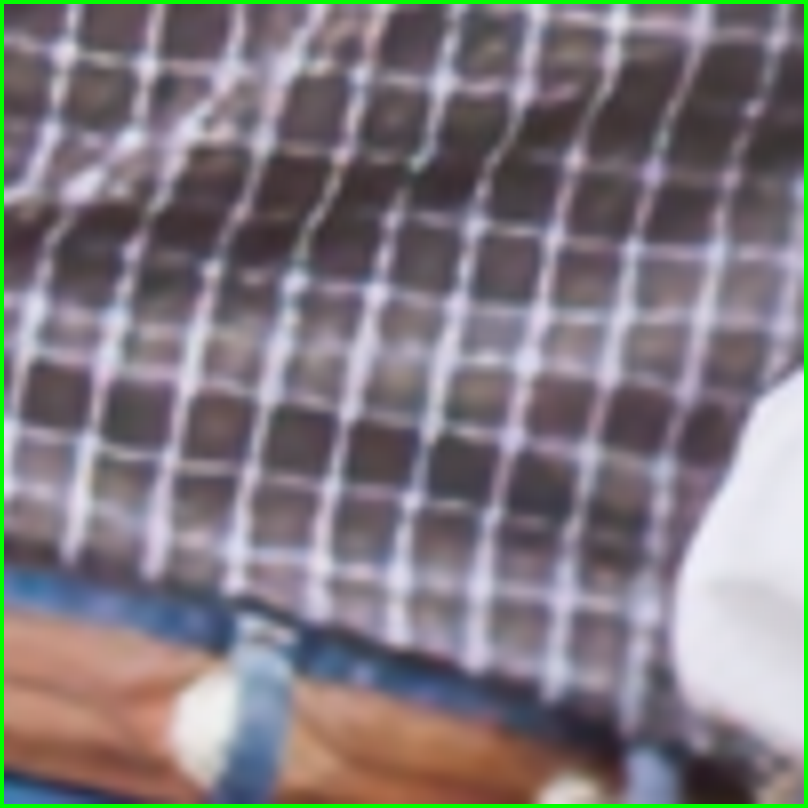}
\end{subfigure} 
&
\begin{subfigure}{\cellasize}
\includegraphics[width=\imgsize]{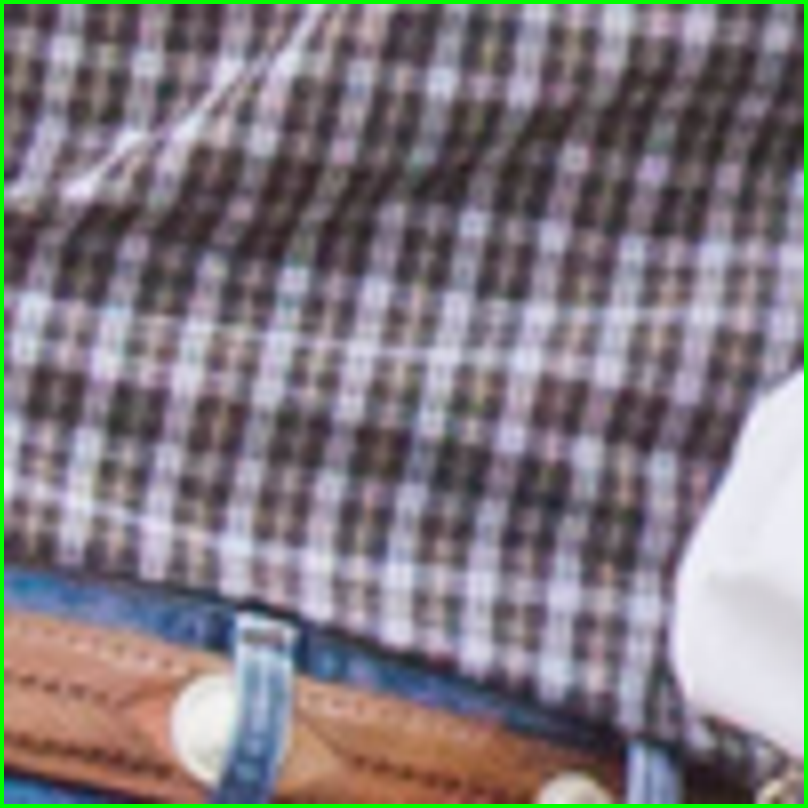}
\end{subfigure}
&
\begin{subfigure}{\cellasize}
\includegraphics[width=\imgsize]{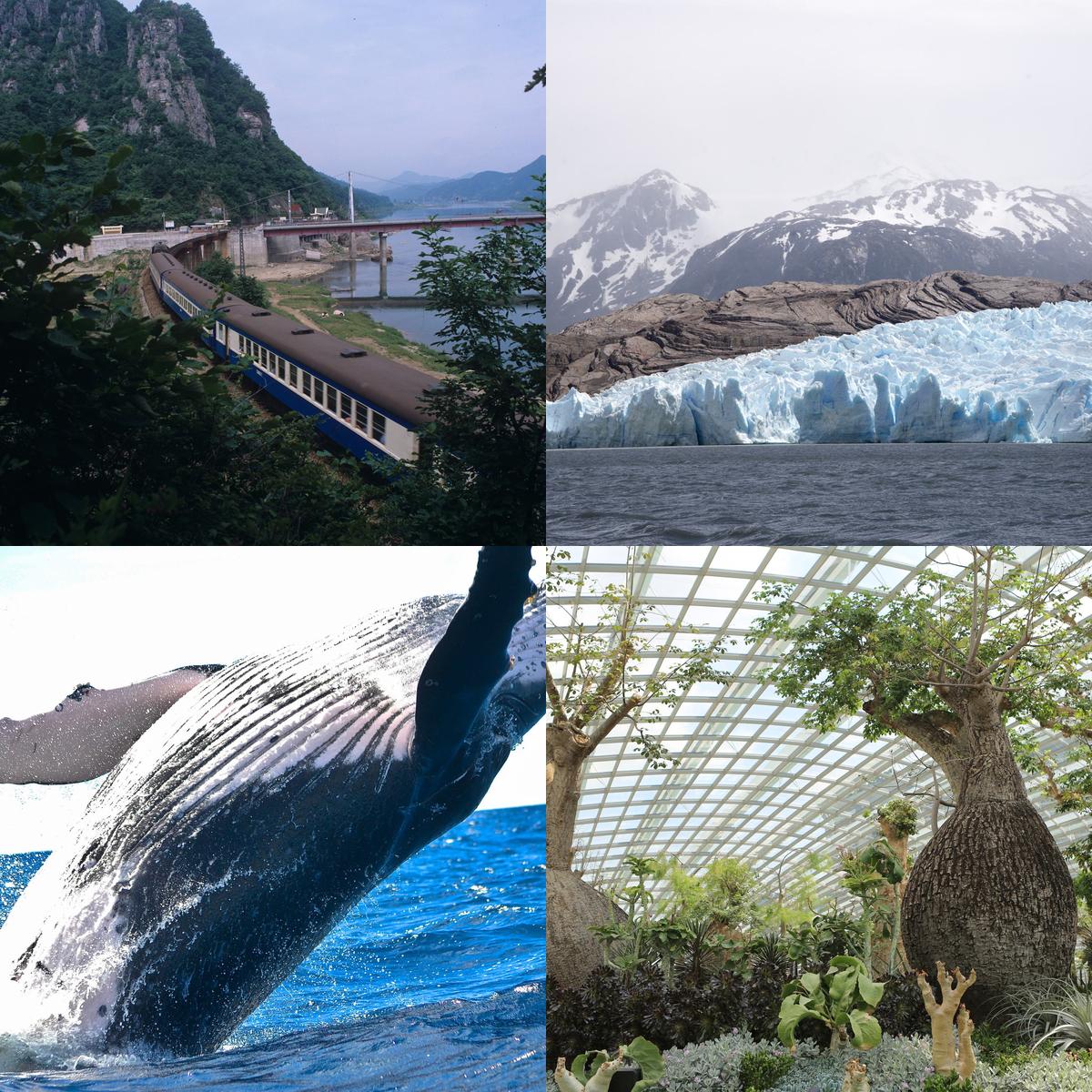}
\end{subfigure}
\\[\rowspace]

\begin{subfigure}{\cellasize}
\includegraphics[width=\imgsize]{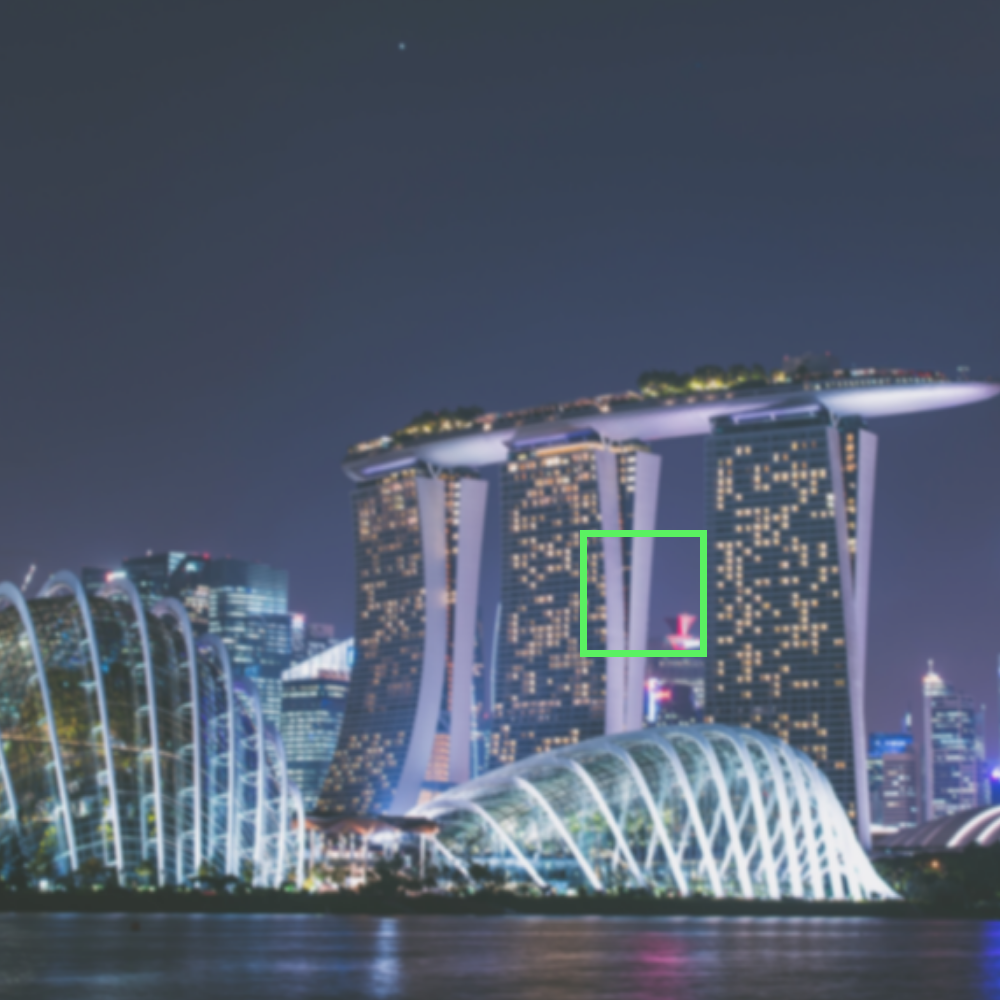}
\end{subfigure}
&
\begin{subfigure}{\cellasize}
\includegraphics[width=\imgsize]{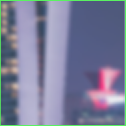}
\end{subfigure}
&
\begin{subfigure}{\cellasize}
\includegraphics[width=\imgsize]{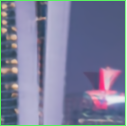}
\end{subfigure} 
&
\begin{subfigure}{\cellasize}
\includegraphics[width=\imgsize]{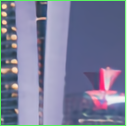}
\end{subfigure}
&
\begin{subfigure}{\cellasize}
\includegraphics[width=\imgsize]{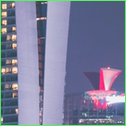}
\end{subfigure}
&
\begin{subfigure}{\cellasize}
\includegraphics[width=\imgsize ]{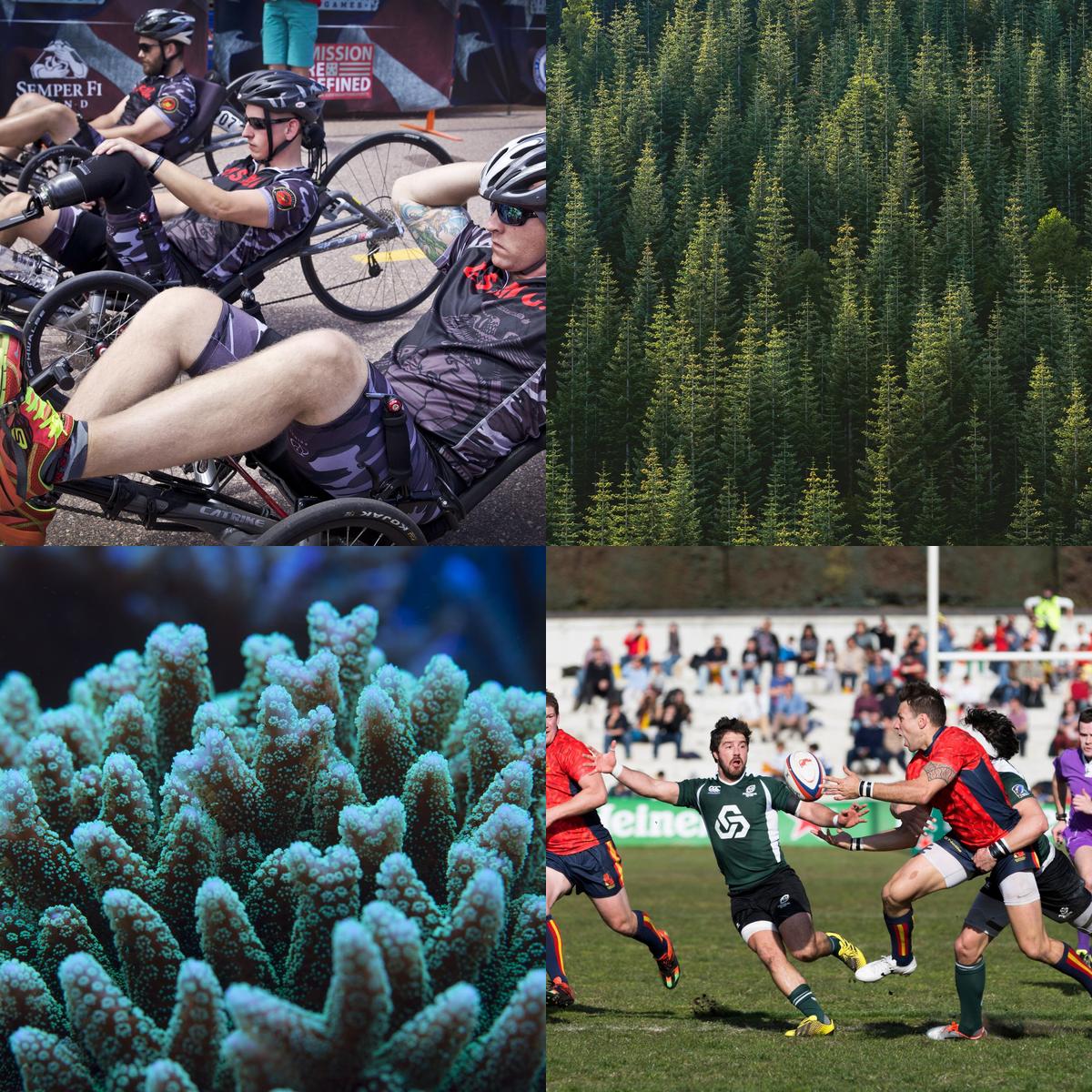}
\end{subfigure}
\\[\rowspace]

\begin{subfigure}{\cellasize}
\includegraphics[width=\imgsize]{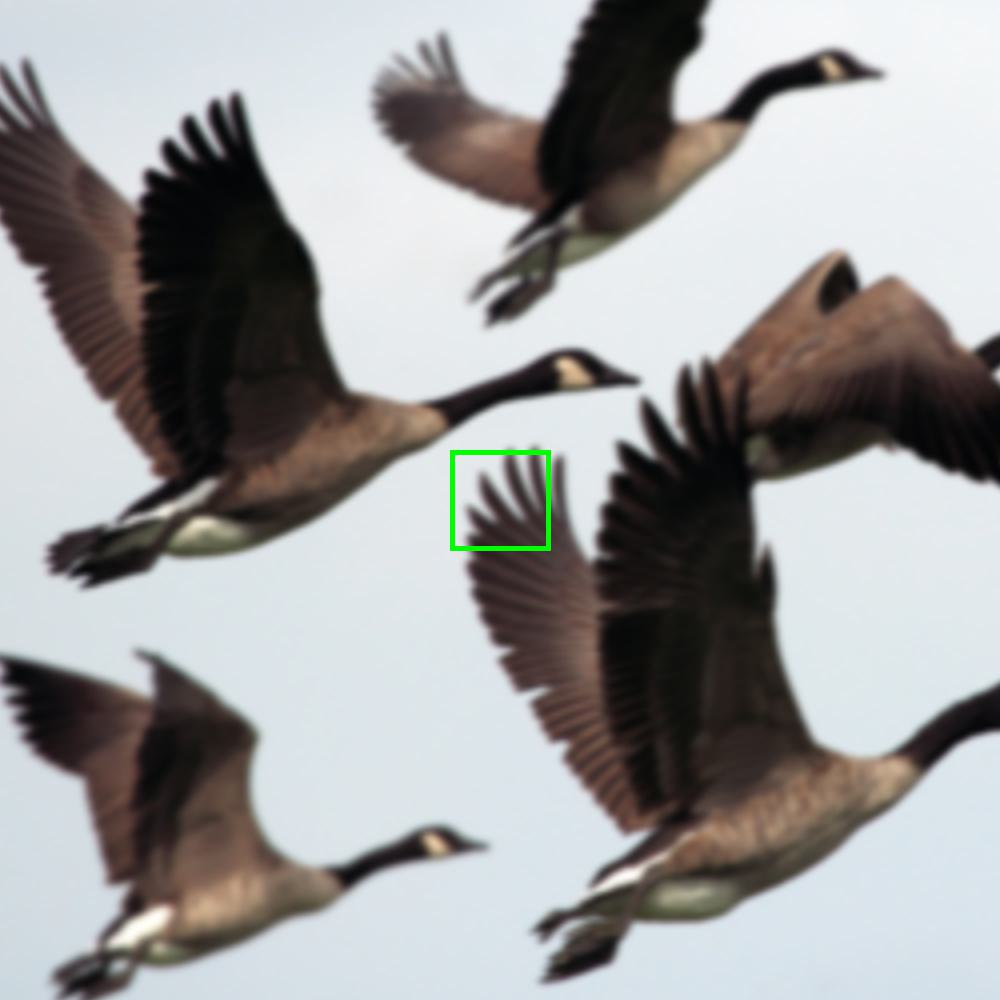}
\end{subfigure}
&
\begin{subfigure}{\cellasize}
\includegraphics[width=\imgsize]{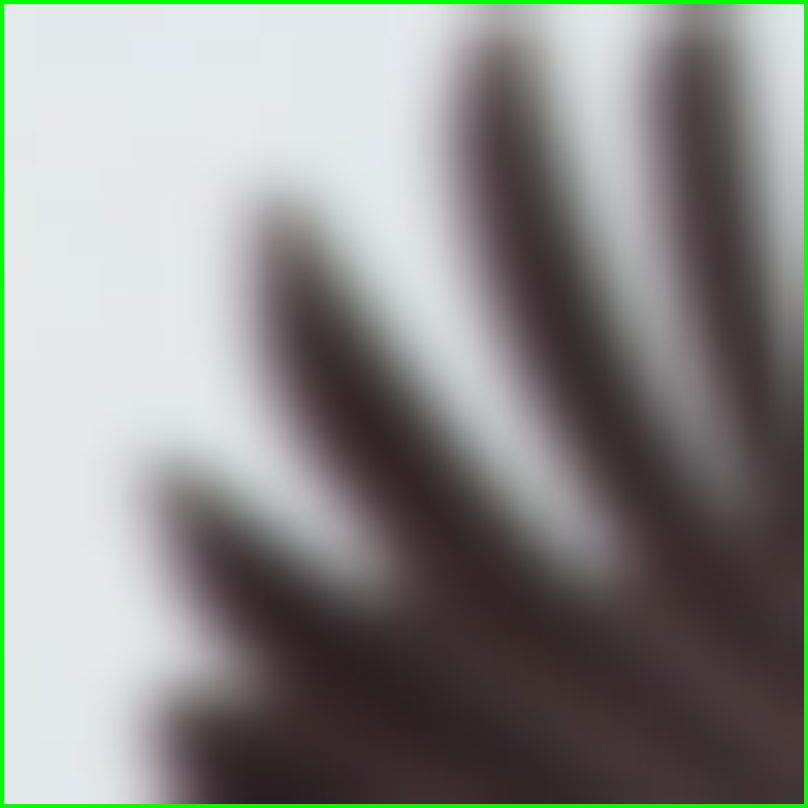}
\end{subfigure}
&
\begin{subfigure}{\cellasize}
\includegraphics[width=\imgsize]{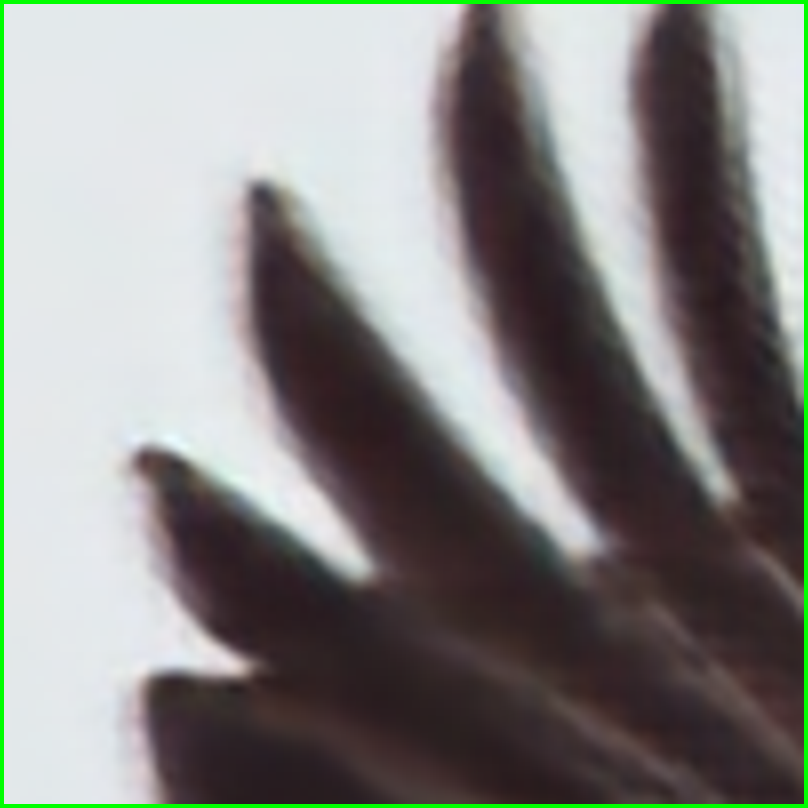}
\end{subfigure}
&
\begin{subfigure}{\cellasize}
\includegraphics[width=\imgsize]{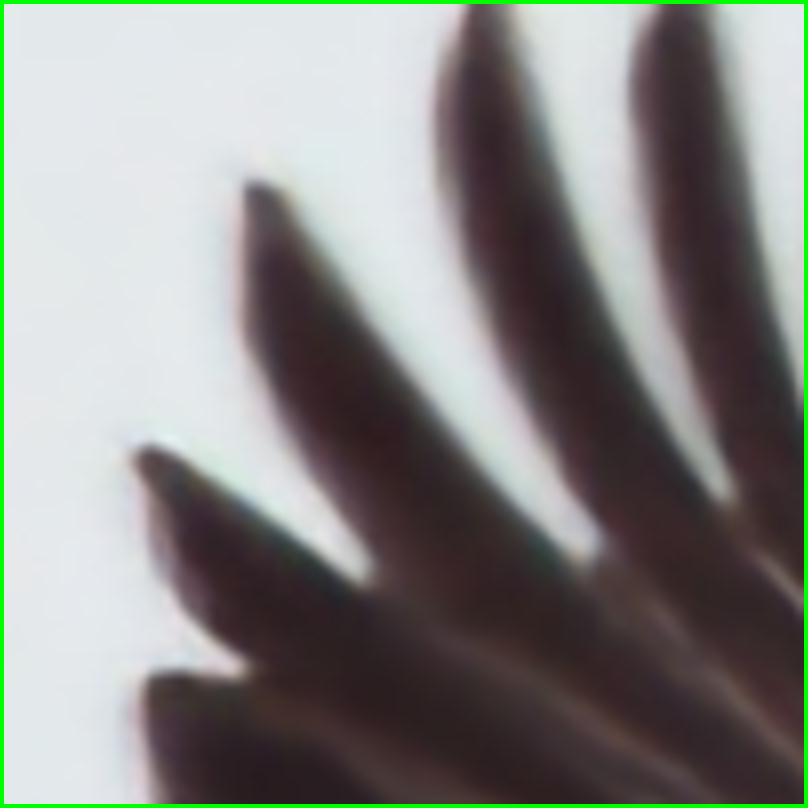}
\end{subfigure}
&
\begin{subfigure}{\cellasize}
\includegraphics[width=\imgsize]{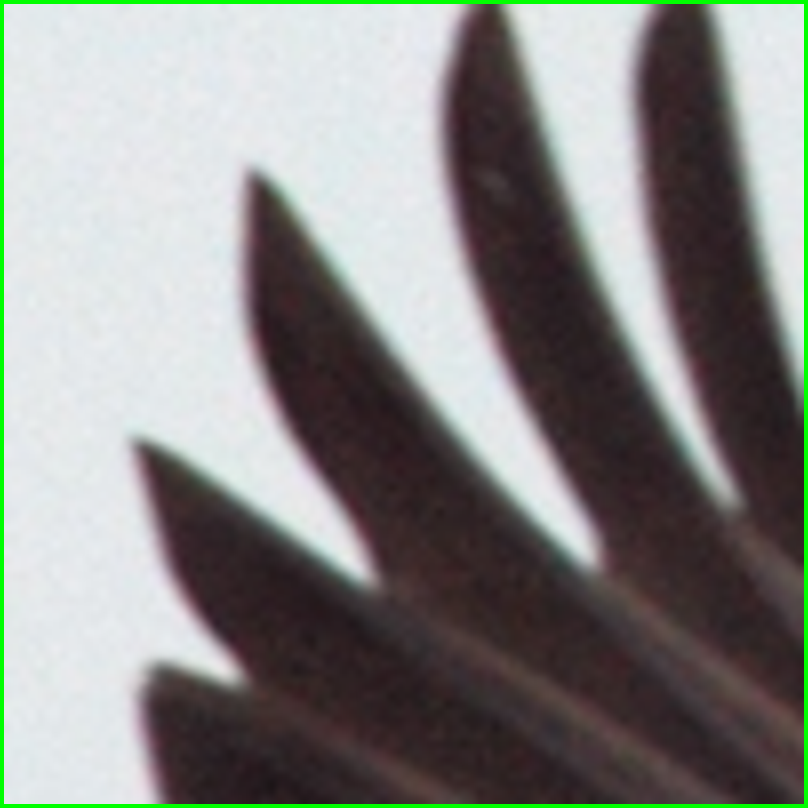}
\end{subfigure}
&
\begin{subfigure}{\cellasize}
\includegraphics[width=\imgsize]{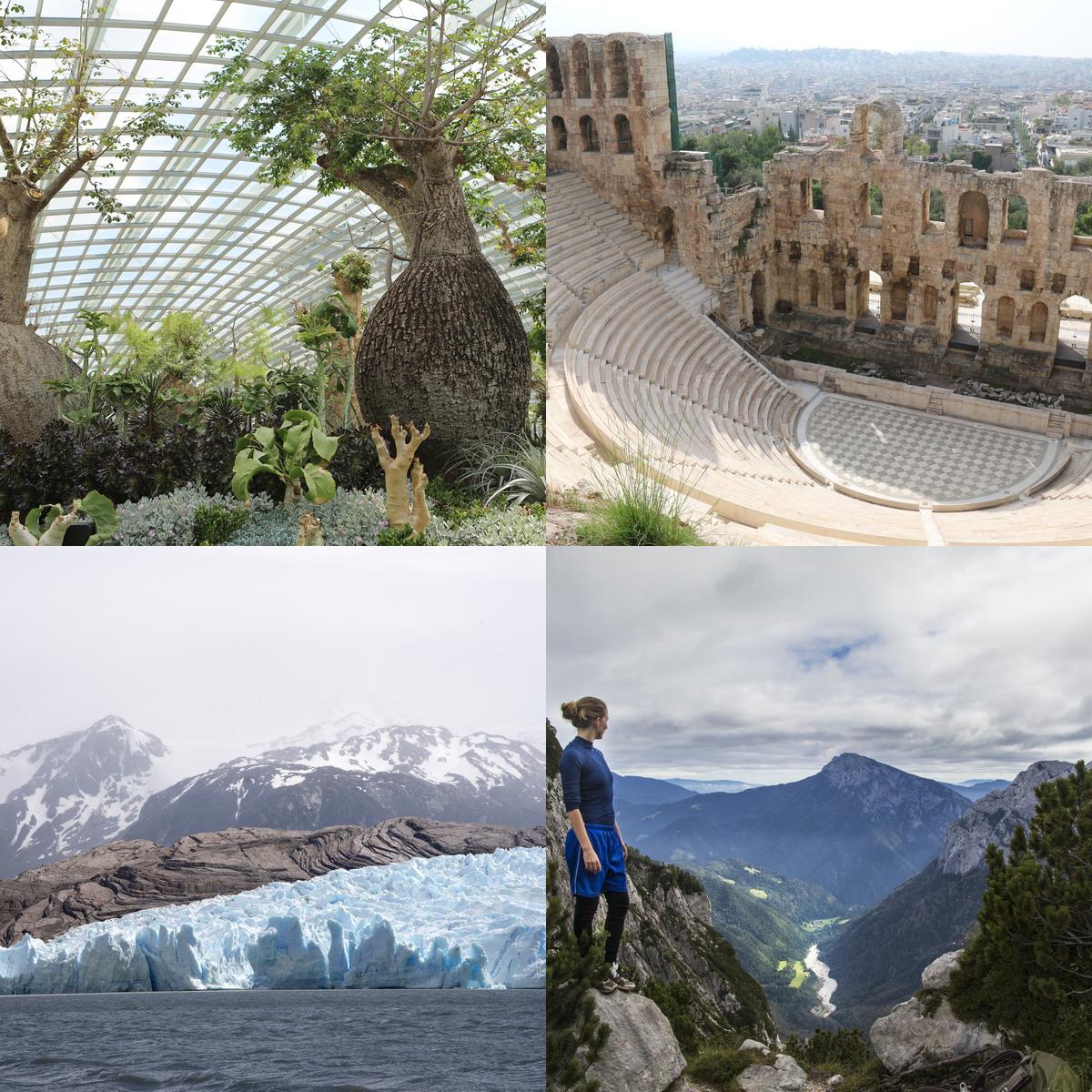}
\end{subfigure}
\\[\rowspace]

\begin{subfigure}{\cellasize}
\includegraphics[width=\imgsize]{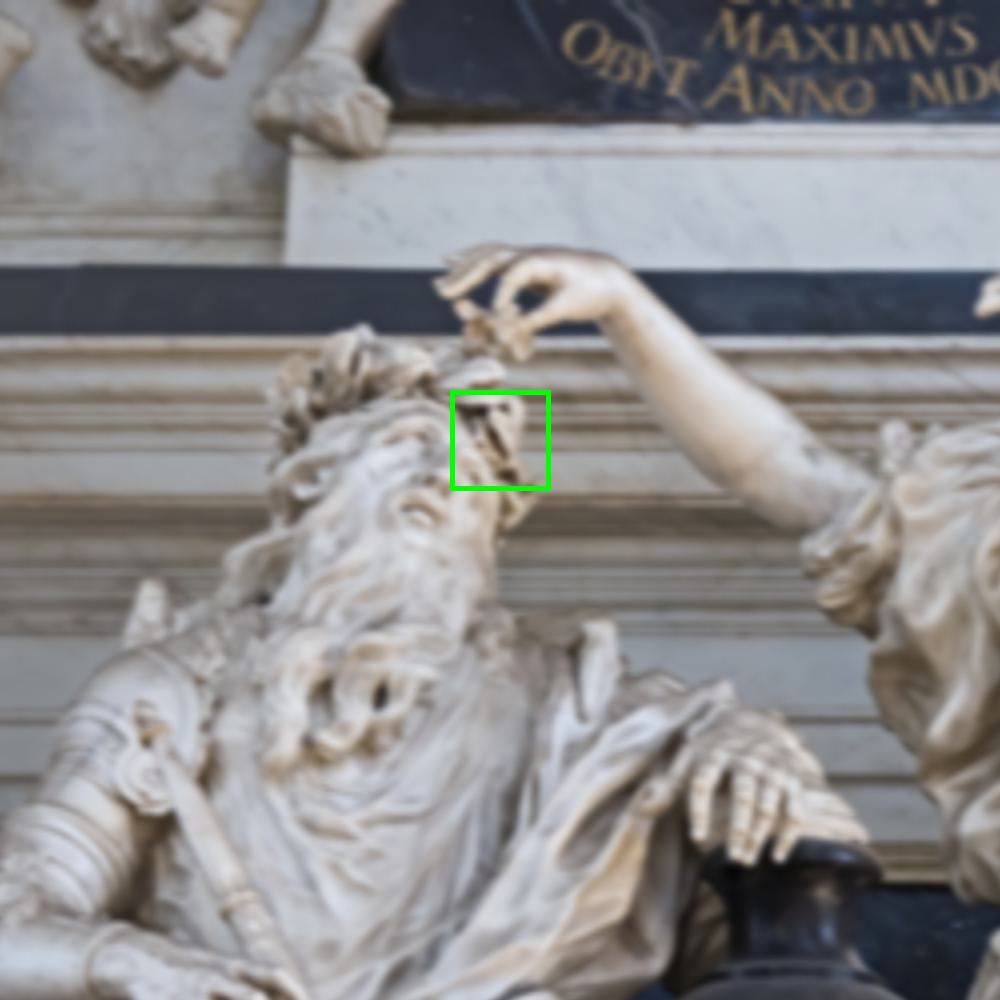}
\end{subfigure}
&
\begin{subfigure}{\cellasize}
\includegraphics[width=\imgsize]{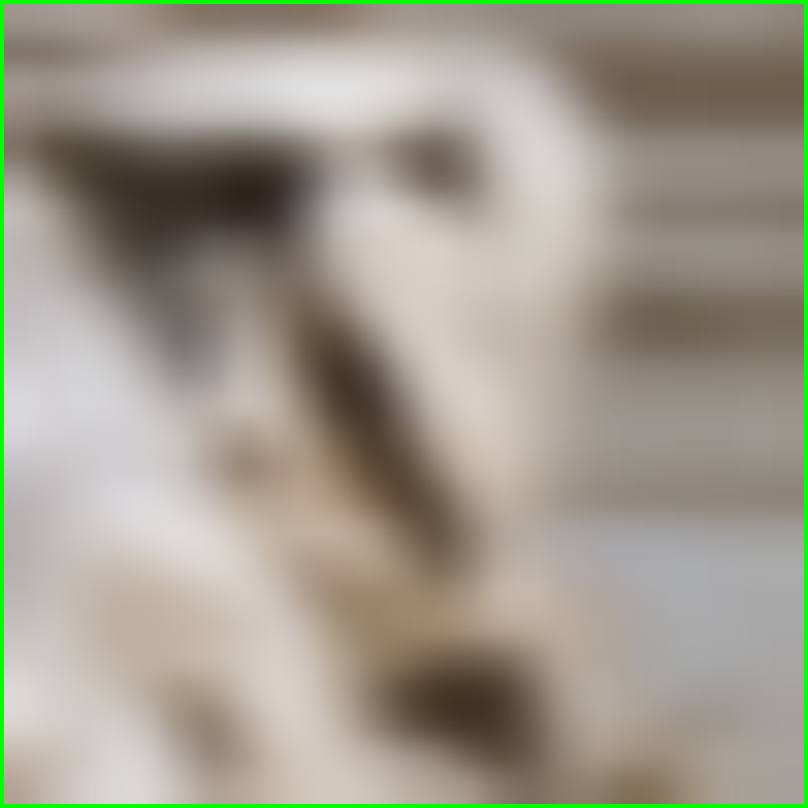}
\end{subfigure}
&
\begin{subfigure}{\cellasize}
\includegraphics[width=\imgsize]{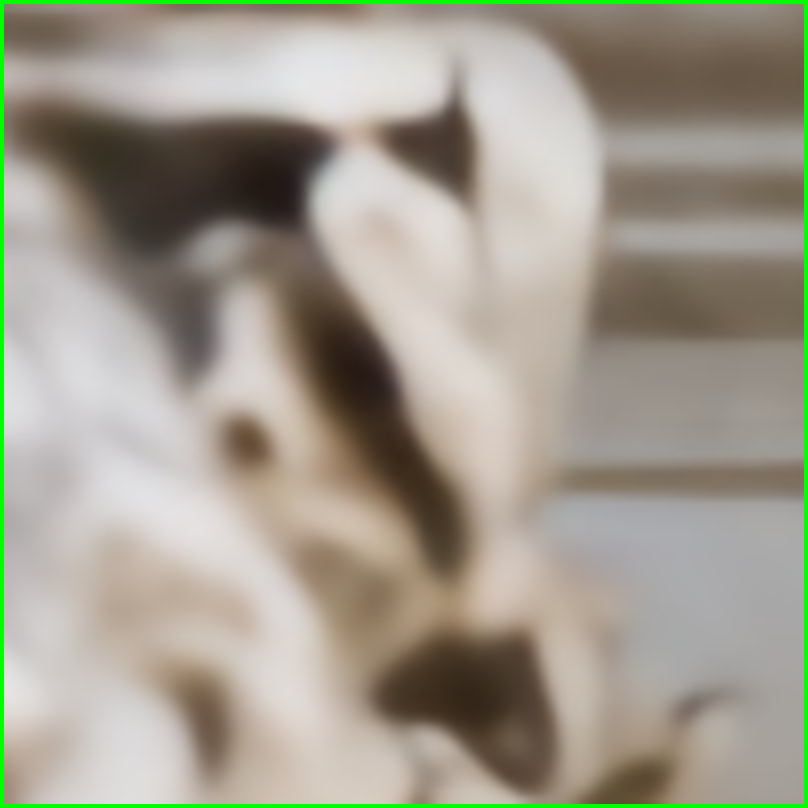}
\end{subfigure}
&
\begin{subfigure}{\cellasize}
\includegraphics[width=\imgsize]{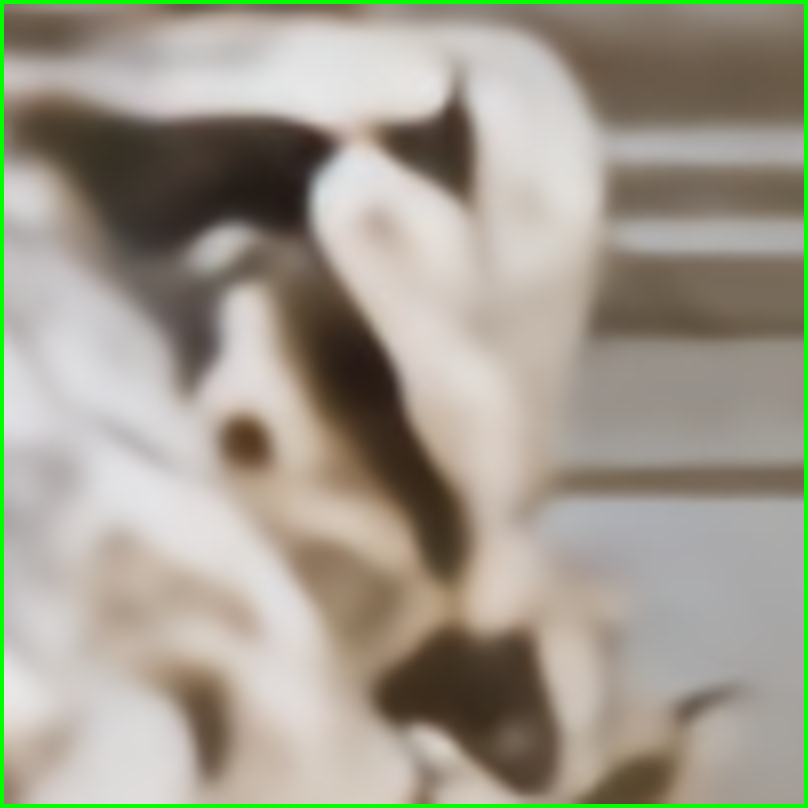}
\end{subfigure}
&
\begin{subfigure}{\cellasize}
\includegraphics[width=\imgsize]{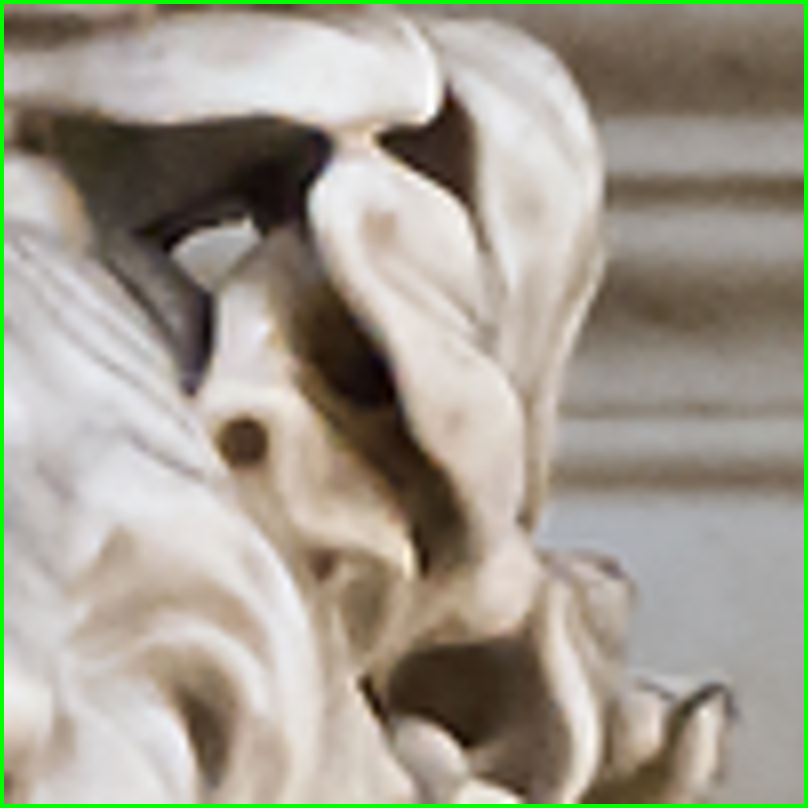}
\end{subfigure}
&
\begin{subfigure}{\cellasize}
\includegraphics[width=\imgsize]{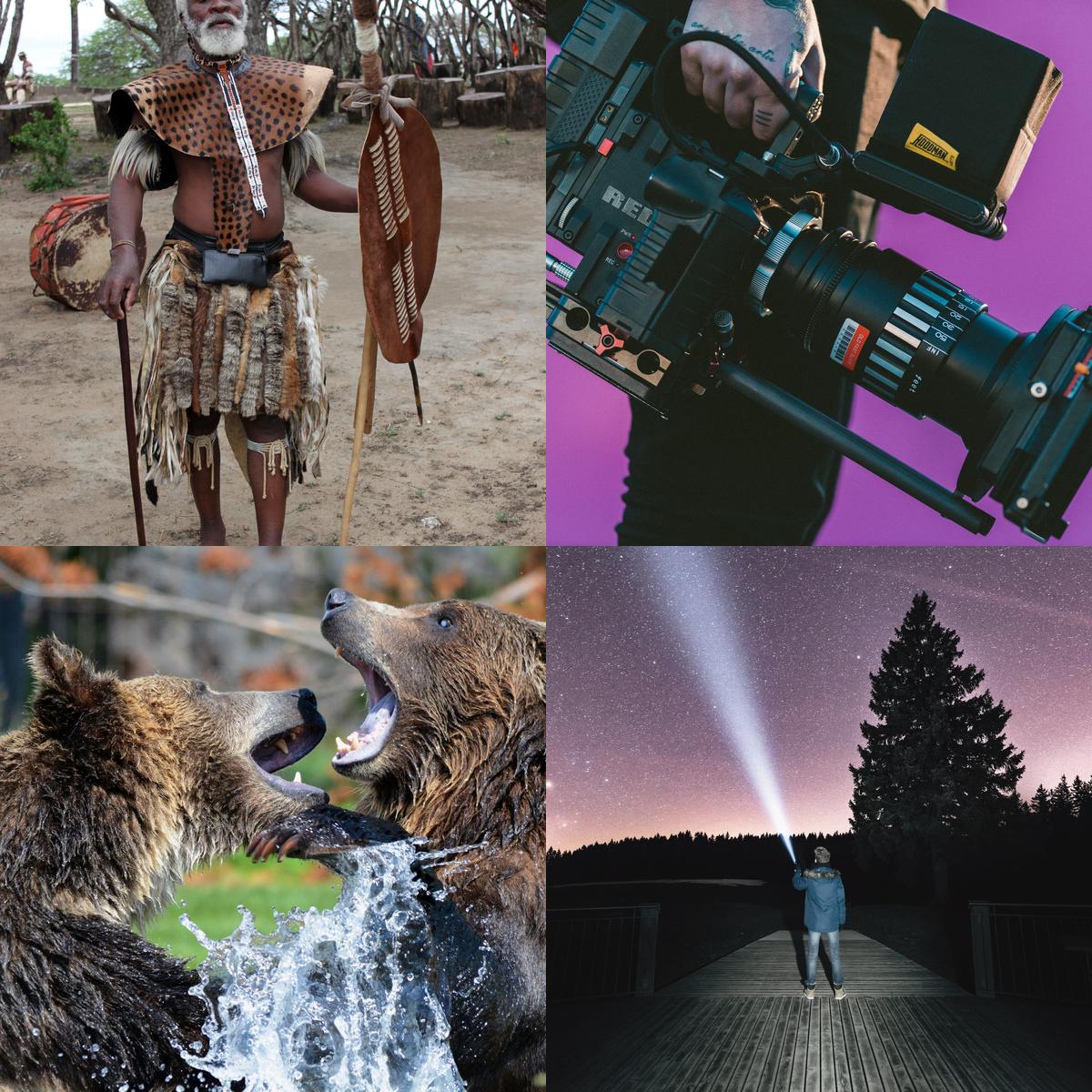}
\end{subfigure}
\\[\rowspace]

\end{tabular}
}
\endgroup
\end{center}

\caption{Qualitative comparison of the RDSR based on "Blind Super Resolution" method. The order from top to bottom sequentially present the results of isotropic and anisotropic kernels at $\times$2 and $\times$4 scaling factors, respectively.
}
\label{fig:vision-BSR-based1}
\end{figure}

\newcommand\tcellsizea{0.33\textwidth}
\newcommand\tcellsizeb{0.1\textwidth}
\newcommand\tcellsizec{0.1\textwidth}

\begin{table}[!htbp]
    \caption{Ablation Study with RDSR x2 with isotropic kernel on DIV2KRK.}
    \label{tab:loss-compare}
    \centering
    \scalebox{0.85}{
    \begin{tabular}{m{\tcellsizea} | m{\tcellsizeb} m{\tcellsizeb} m{\tcellsizeb} | m{\tcellsizec} m{\tcellsizec}}
    \toprule
    \multicolumn{1}{m{\tcellsizea}|} {\centering Method} &     
    \multicolumn{1}{m{\tcellsizeb}}{\centering $\mathcal{L}_{ref}$} &
    \multicolumn{1}{m{\tcellsizeb}}{\centering $\mathcal{L}_{gan}$} &
    \multicolumn{1}{m{\tcellsizeb}|}{\centering $\mathcal{L}_{reg}$} &
    \multicolumn{1}{m{\tcellsizec}}{\centering PSNR} &
    \multicolumn{1}{m{\tcellsizec}}{\centering SSIM} \\
    \midrule
    \multicolumn{1}{m{\tcellsizea}|} {DASR $iso$} &     
    \multicolumn{1}{m{\tcellsizeb}}{\centering \xmark} &
    \multicolumn{1}{m{\tcellsizeb}}{\centering \xmark} &
    \multicolumn{1}{m{\tcellsizeb}|}{\centering \xmark} &
    \multicolumn{1}{m{\tcellsizec}}{\centering 30.743} &
    \multicolumn{1}{m{\tcellsizec}}{\centering 0.8586} \\
    
    \multicolumn{1}{m{\tcellsizea}|} {RDSR $iso$ $(baseline)$} &
    \multicolumn{1}{m{\tcellsizeb}}{\centering \checkmark} &
    \multicolumn{1}{m{\tcellsizeb}}{\centering \xmark} &
    \multicolumn{1}{m{\tcellsizeb}|}{\centering \xmark} &
    \multicolumn{1}{m{\tcellsizec}}{\centering 30.876} &
    \multicolumn{1}{m{\tcellsizec}}{\centering 0.8609} \\

    \multicolumn{1}{m{\tcellsizea}|} {RDSR $iso$ + $\mathcal{L}_{gan}$} &     
    \multicolumn{1}{m{\tcellsizeb}}{\centering \checkmark} &
    \multicolumn{1}{m{\tcellsizeb}}{\centering \checkmark} &
    \multicolumn{1}{m{\tcellsizeb}|}{\centering \xmark} &
    \multicolumn{1}{m{\tcellsizec}}{\centering 30.927} &
    \multicolumn{1}{m{\tcellsizec}}{\centering 0.8621} \\

    \multicolumn{1}{m{\tcellsizea}|} {RDSR $iso$ + $\mathcal{L}_{gan}$ + $\mathcal{L}_{reg}$} &     
    \multicolumn{1}{m{\tcellsizeb}}{\centering \checkmark} &
    \multicolumn{1}{m{\tcellsizeb}}{\centering \checkmark} &
    \multicolumn{1}{m{\tcellsizeb}|}{\centering \checkmark} &
    \multicolumn{1}{m{\tcellsizec}}{\centering \textbf{30.973}} &
    \multicolumn{1}{m{\tcellsizec}}{\centering \textbf{0.8638}} \\
    \bottomrule
    \end{tabular}
    }
\end{table}

\subsection{Ablation Study}
In this section, we study our RDSR model with different loss settings to evaluate their impact. Additionally, we investigate the importance of finding the preliminary degradation process from target LR images during the model initialization phase through experiments with ground truth (GT) kernels. Furthermore, one of our key contributions is the integration of HR images, where the characteristics of HR images—such as their types and quantity—play a critical role in influencing the final outcomes. Therefore, we analyze the impact of HR image selection methods and the number of reference images used in our study.
 
\subsubsection{The Designed Loss Functions.} Table \ref{tab:loss-compare} demonstrates the effectiveness of our designed loss functions. Experimental results on the DIV2K$\times2$ dataset under isotropic settings show that integrating the adversarial loss $L_{gan}$ improved performance by 0.05 dB in PSNR and 0.0012 in SSIM compared to the baseline. Furthermore, the inclusion of the regularization loss $L_{reg}$ resulted in an additional gain of 0.046 dB in PSNR and 0.0017 in SSIM.

\subsubsection{Ground-truth Kernel.} In our architecture, the performance is minimally impacted by the content of the reference images. However, the degradation process has been identified as a critical factor. Table \ref{tab:gt-kernel-compare} illustrates the substantial performance improvements achieved by using the ground truth kernel in place of the downsampling network $G_{dn}$. If a more effective kernel estimation method is discovered, our approach can readily integrate it by substituting the new method into the downsampling network.

\begin{table}[!htbp]
    \caption{Comparison of ground truth kernel on RDSR x2 with anisotropic kernel.}
    \label{tab:gt-kernel-compare}
    \centering
    \scalebox{0.8}{
    \begin{tabular}{m{0.4\linewidth} m{0.15\linewidth} m{0.15\linewidth} m{0.15\linewidth} } \toprule
    Method & scale & PSNR & SSIM \\ \midrule
    DASR $aniso$ & $\times$2 & 31.051 & 0.8668 \\
    RDSR $aniso$ & $\times$2 & 31.161 & 0.8683 \\
    RDSR $aniso$ + GT kernel & $\times$2 & \textbf{33.144} & \textbf{0.8994}  \\\bottomrule
    \end{tabular}
    }
\end{table}

\begin{table}[!htbp]
    \caption{Comparison of reference image selection policies on DIV2K x2 datasets.}
     
    \label{tab:ref-selection}
    \centering
    \scalebox{0.8}{
    \begin{tabular}{m{0.30\linewidth} m{0.30\linewidth}  m{0.12\linewidth} m{0.12\linewidth} } \toprule
    Method \& Datasets & Ref. Selection Method & PSNR-Y & SSIM \\ \midrule
    \multirow{3}*{\makecell{DualSR + Ref. add-on \\ DIV2K x2}} & auto-selection & \textbf{26.228} & \textbf{0.7353}  \\
    & random selection & 26.190 & 0.7340  \\
    & reverse auto-selection & 26.164 & 0.7338  \\\bottomrule
    \end{tabular}}
\end{table}

\begin{figure}[!htbp]
\centering
\small
\begingroup
\setlength{\tabcolsep}{0.2pt} 
\renewcommand{\arraystretch}{1} 
\scalebox{0.85}{
\begin{tabular}{ m{\cellsize} m{\cellsize} m{\cellsize} m{\cellsize} m{\cellsize} m{\cellsize}}
LR \centering & Bicubic \centering & DualSR \centering & RDSR(ours) & Ground Truth & \hspace{1.7em} Ref. \\

\begin{subfigure}{\cellasize}
\includegraphics[width=\imgsize]{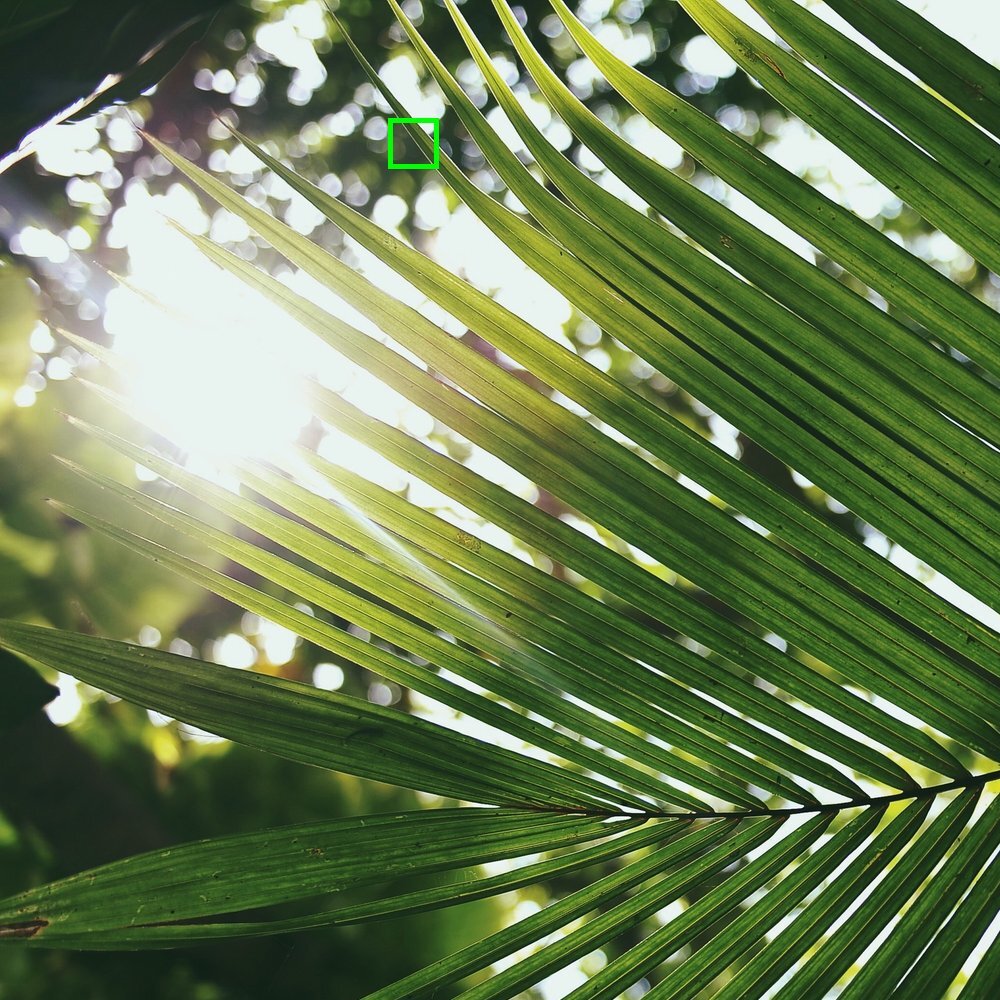}
\end{subfigure} 
&
\begin{subfigure}{\cellasize}
\includegraphics[width=\imgsize]{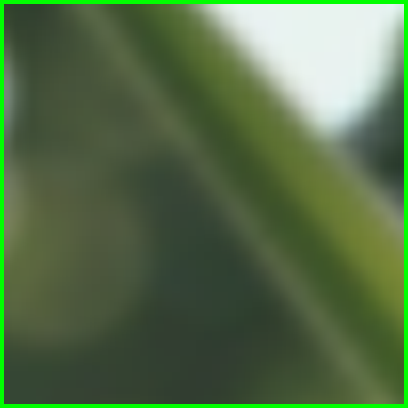}
\end{subfigure} 
&
\begin{subfigure}{\cellasize}
\includegraphics[width=\imgsize]{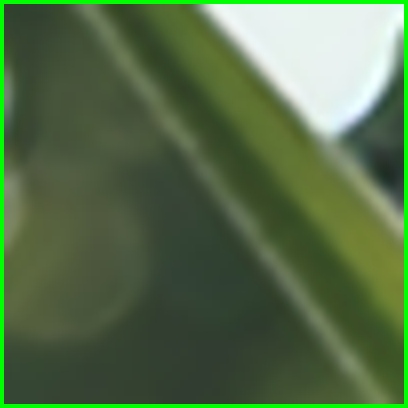}
\end{subfigure} 
& 
\begin{subfigure}{\cellasize}
\includegraphics[width=\imgsize]{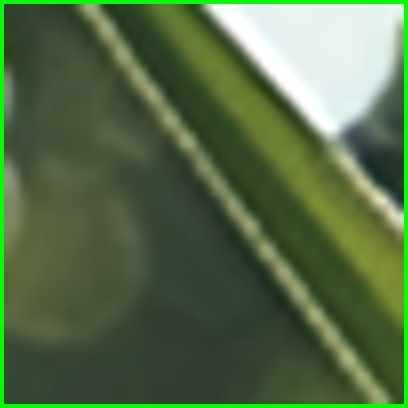}
\end{subfigure} 
&
\begin{subfigure}{\cellasize}
\includegraphics[width=\imgsize]{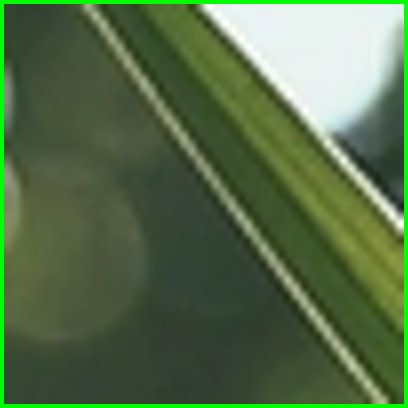}
\end{subfigure}
&
\begin{subfigure}{\cellasize}
\includegraphics[width=\imgsize]{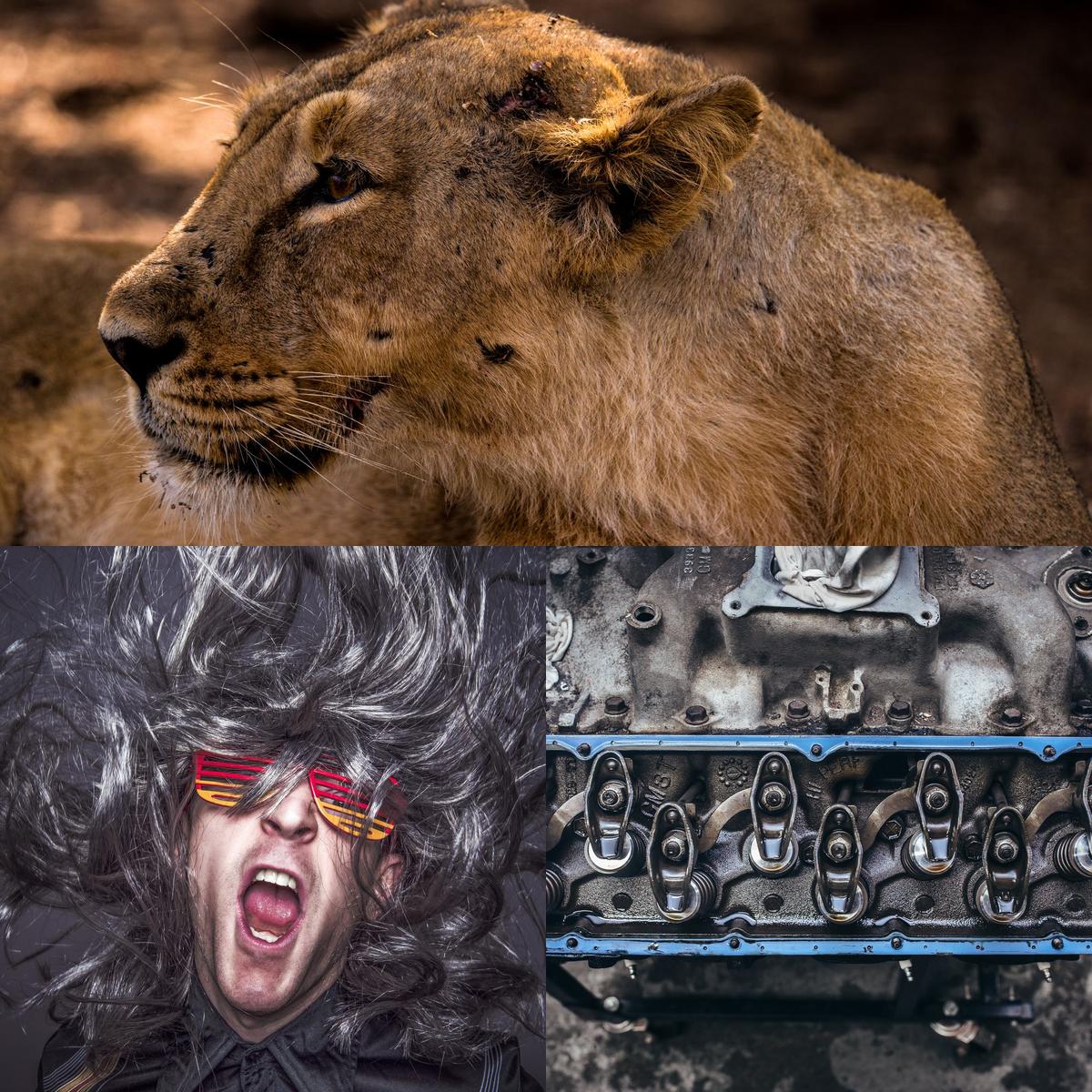}
\end{subfigure}
\\[\rowspace]

\begin{subfigure}{\cellasize}
\includegraphics[width=\imgsize]{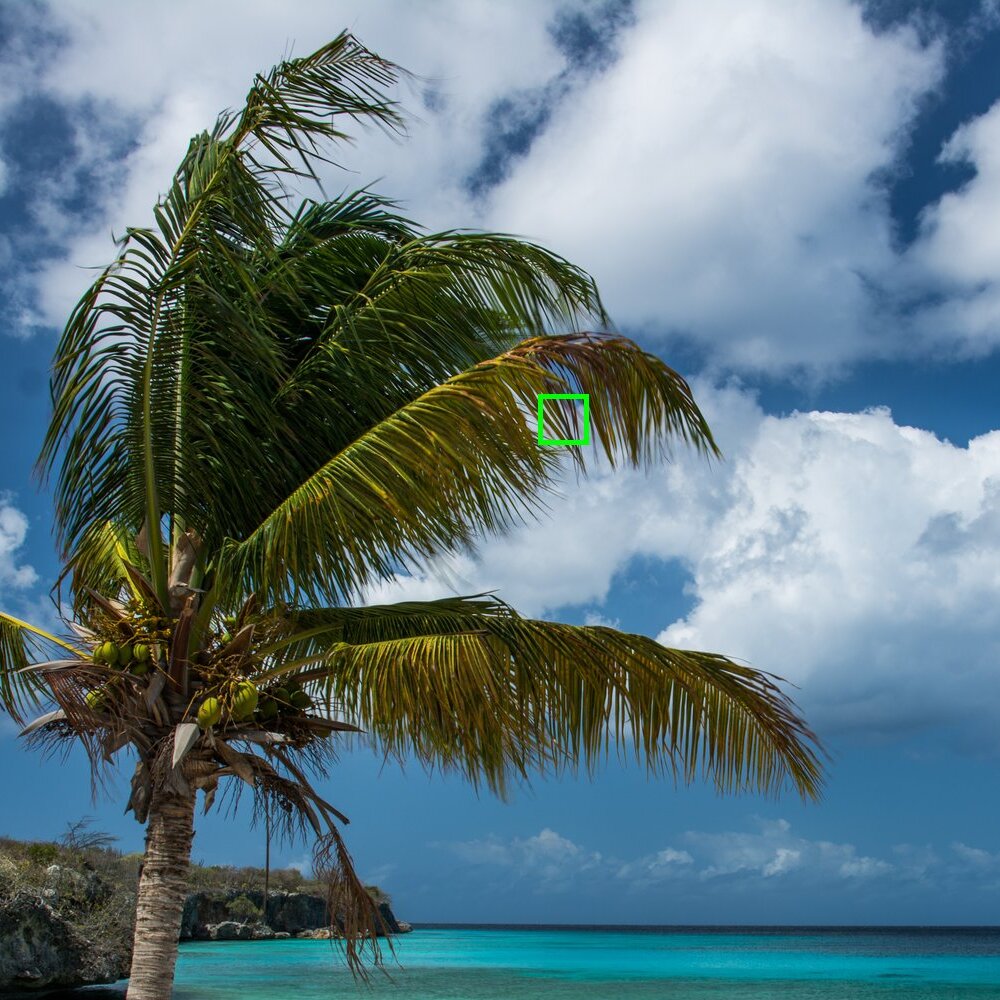}
\end{subfigure}
&

\begin{subfigure}{\cellasize}
\includegraphics[width=\imgsize]{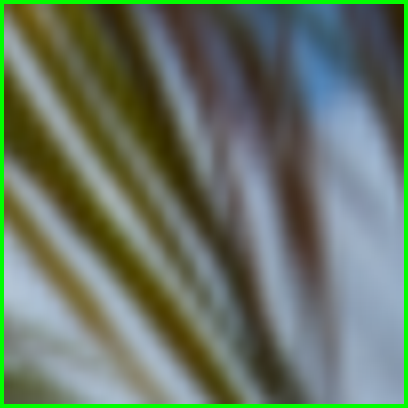}
\end{subfigure}
&
\begin{subfigure}{\cellasize}
\includegraphics[width=\imgsize]{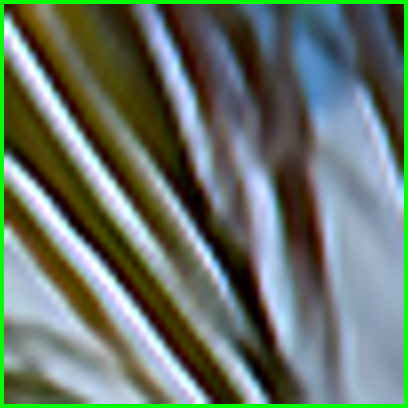}
\end{subfigure} 
&
\begin{subfigure}{\cellasize}
\includegraphics[width=\imgsize]{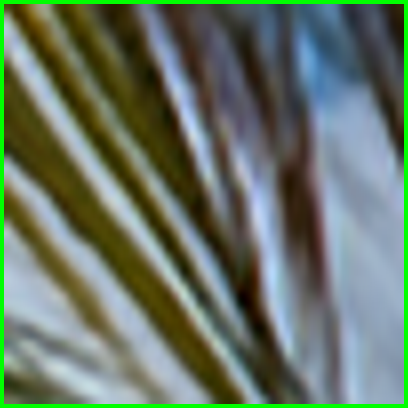}
\end{subfigure}
&
\begin{subfigure}{\cellasize}
\includegraphics[width=\imgsize]{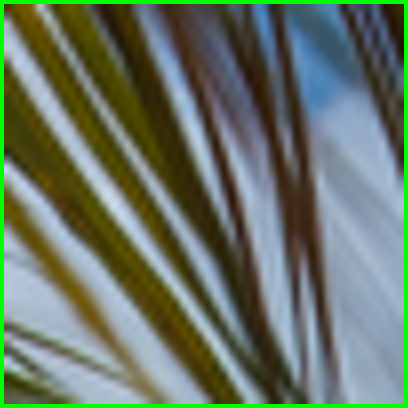}
\end{subfigure}
&
\begin{subfigure}{\cellasize}
\includegraphics[width=\imgsize]{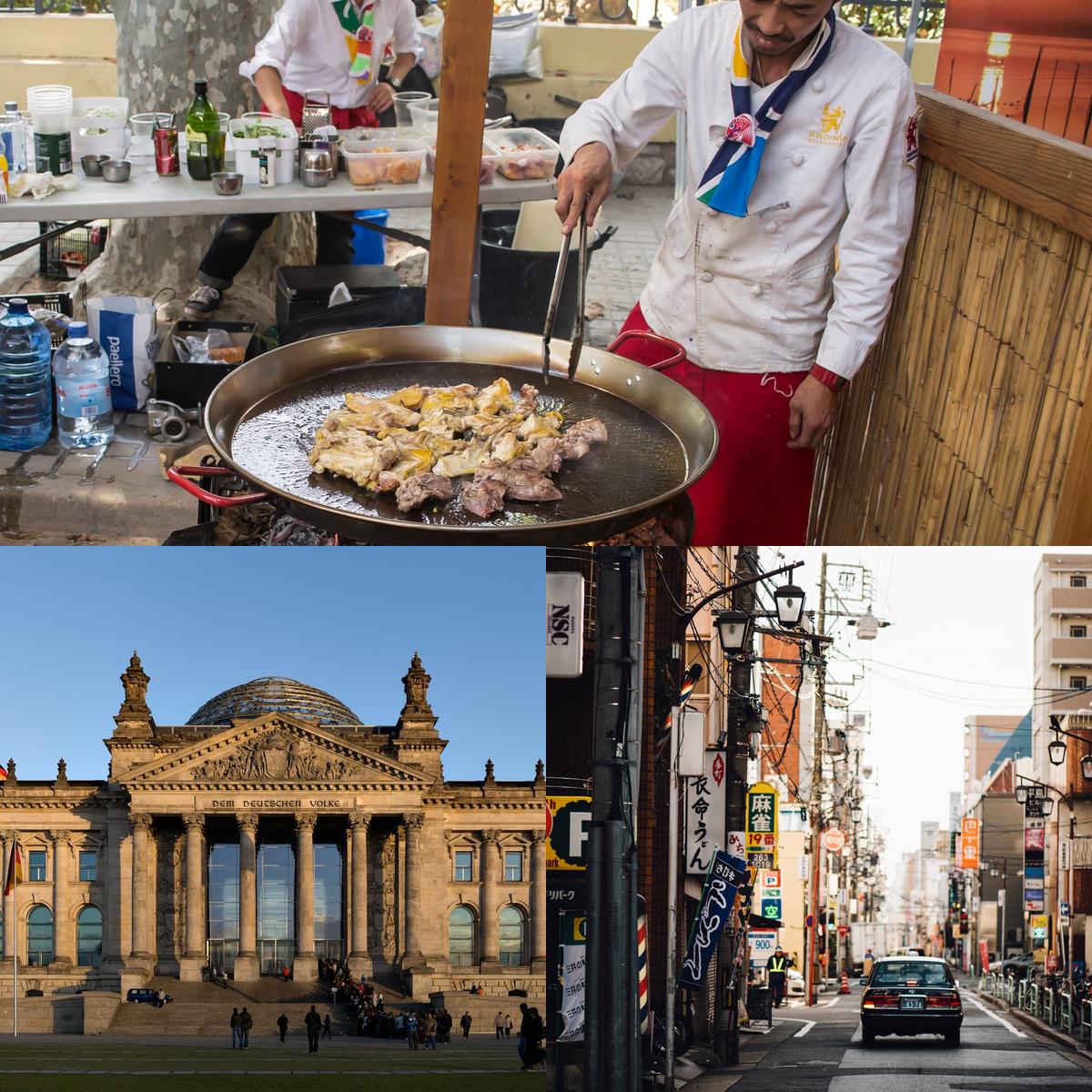}
\end{subfigure}
\\[\rowspace] \midrule

\begin{subfigure}{\cellasize}
\includegraphics[width=\imgsize]{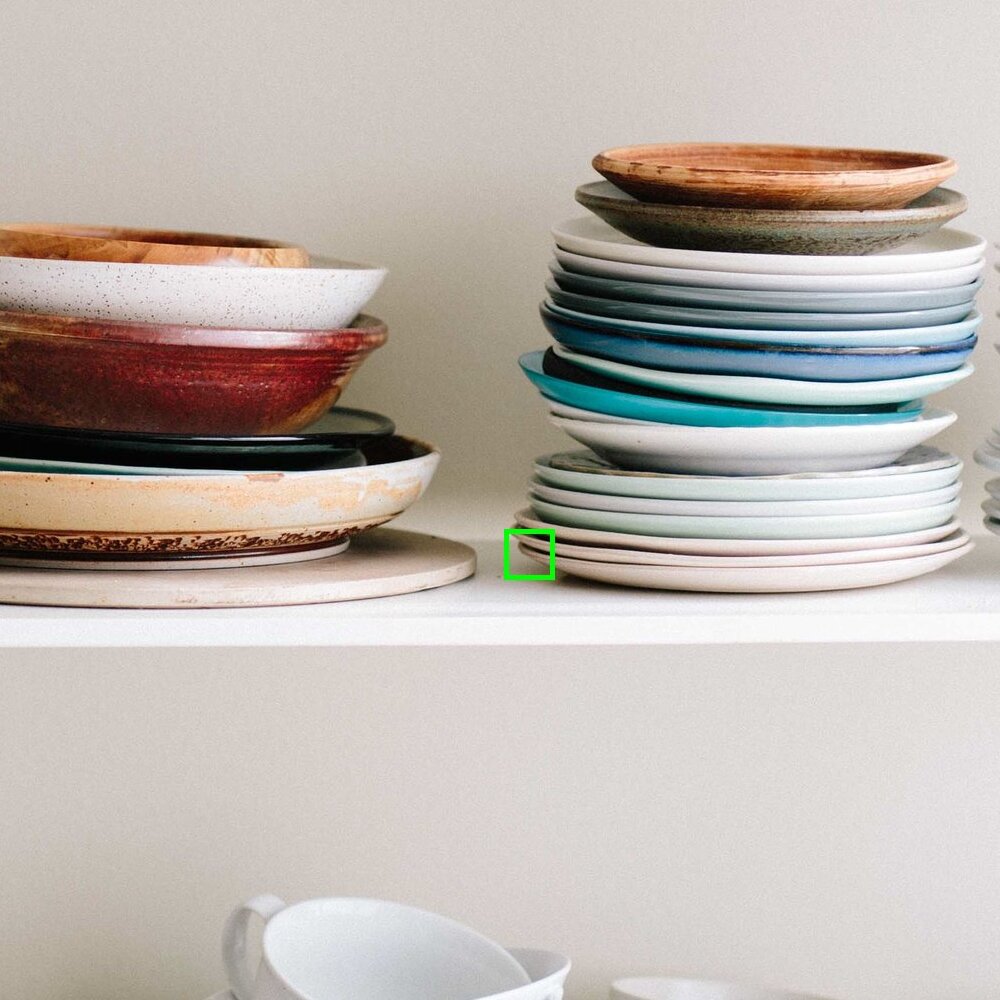}
\end{subfigure}
&
\begin{subfigure}{\cellasize}
\includegraphics[width=\imgsize]{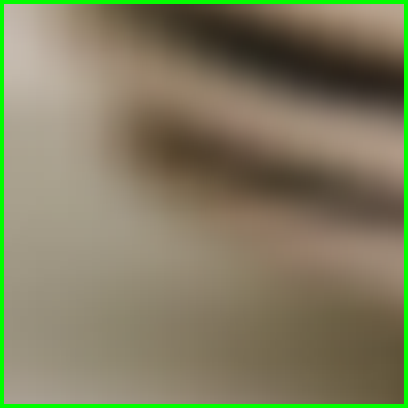}
\end{subfigure}
&
\begin{subfigure}{\cellasize}
\includegraphics[width=\imgsize]{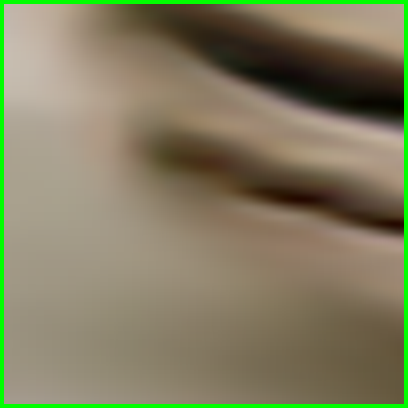}
\end{subfigure}
&
\begin{subfigure}{\cellasize}
\includegraphics[width=\imgsize]{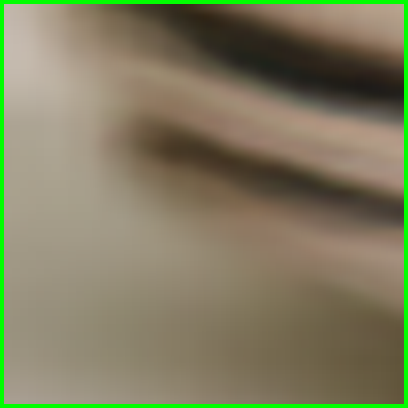}
\end{subfigure}
&
\begin{subfigure}{\cellasize}
\includegraphics[width=\imgsize]{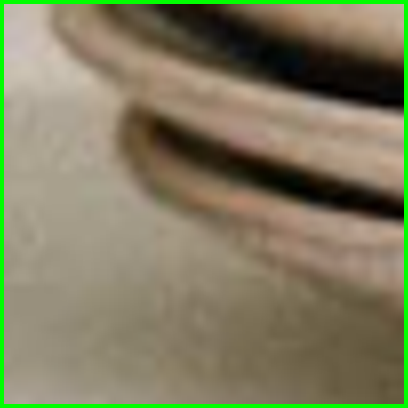}
\end{subfigure}
&
\begin{subfigure}{\cellasize}
\includegraphics[width=\imgsize]{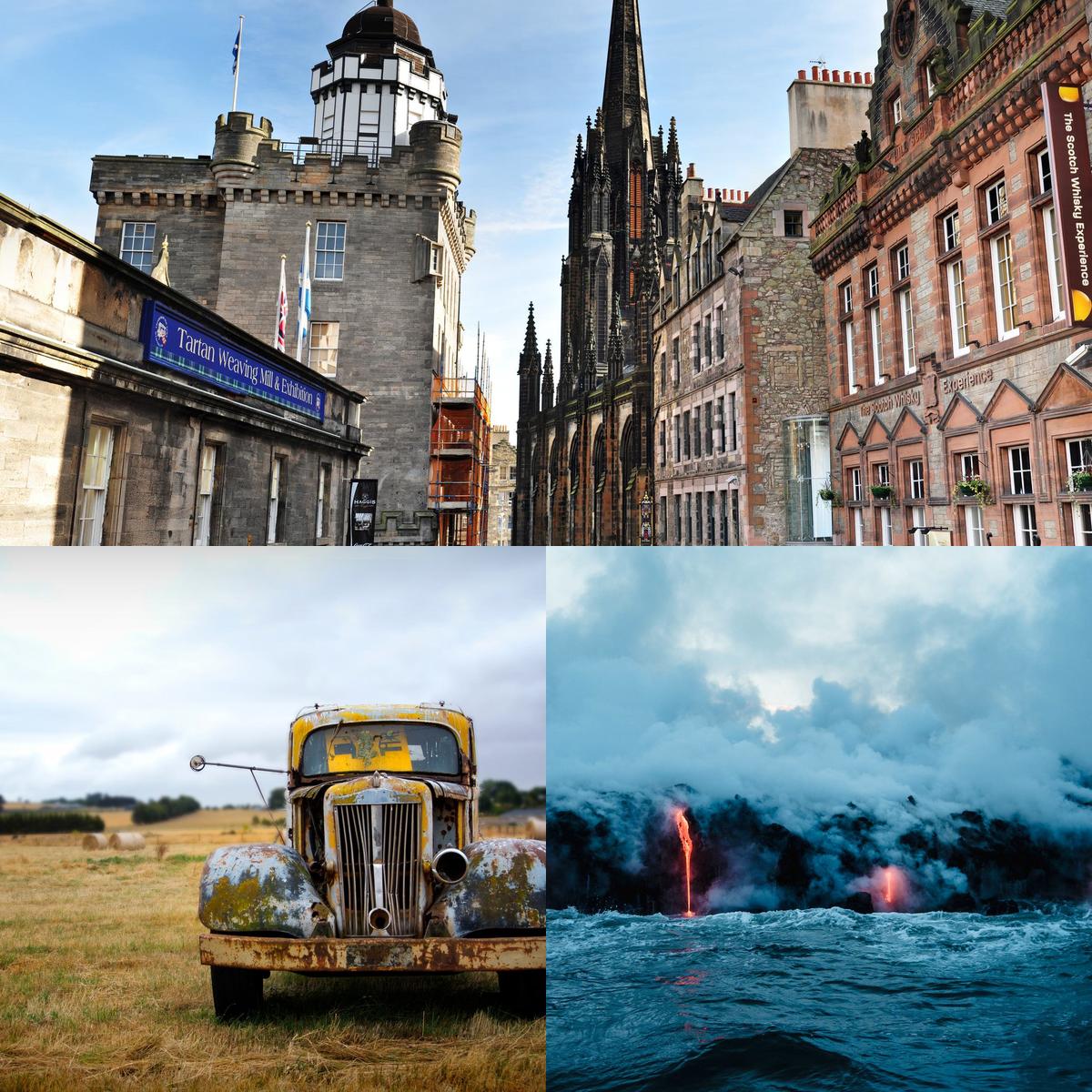}
\end{subfigure}
\\[\rowspace]

\begin{subfigure}{\cellasize}
\includegraphics[width=\imgsize]{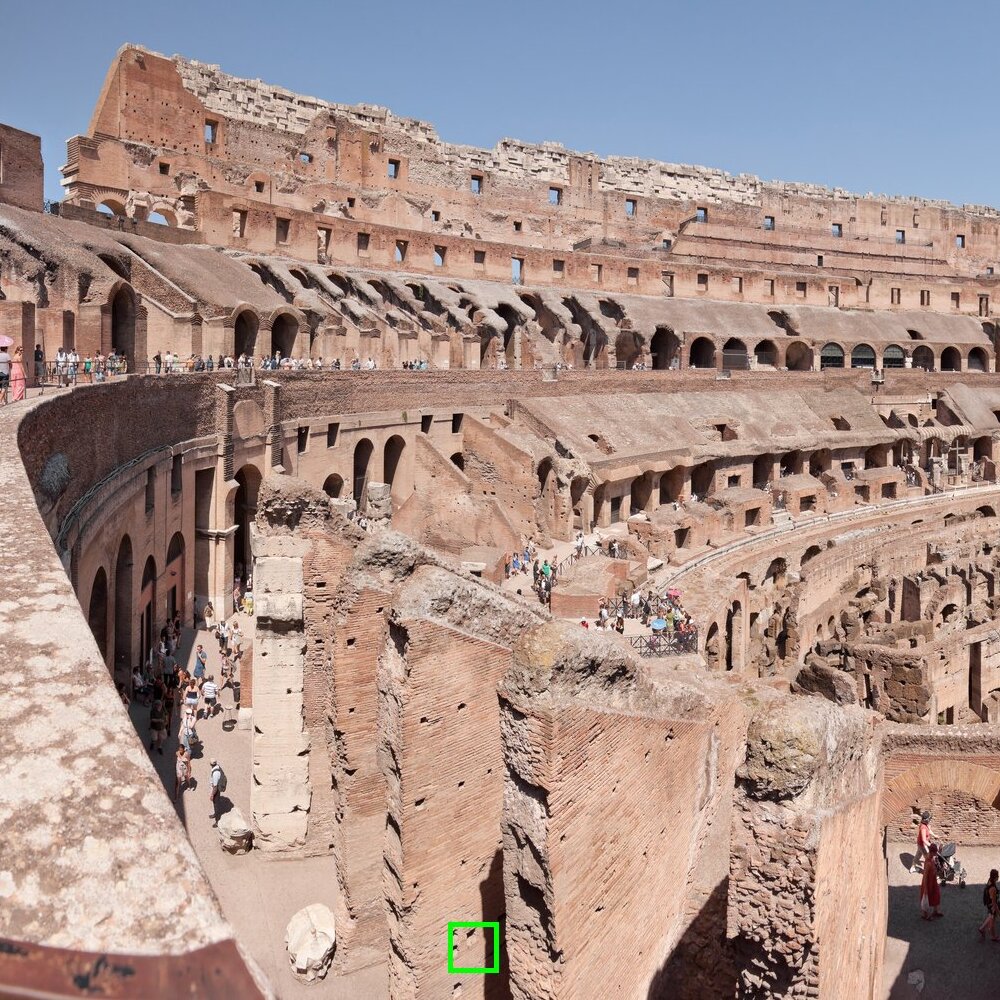}
\end{subfigure}
&
\begin{subfigure}{\cellasize}
\includegraphics[width=\imgsize]{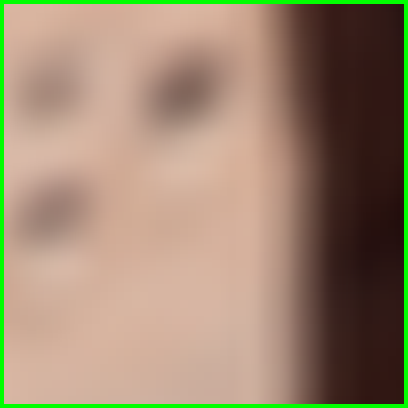}
\end{subfigure}
&
\begin{subfigure}{\cellasize}
\includegraphics[width=\imgsize]{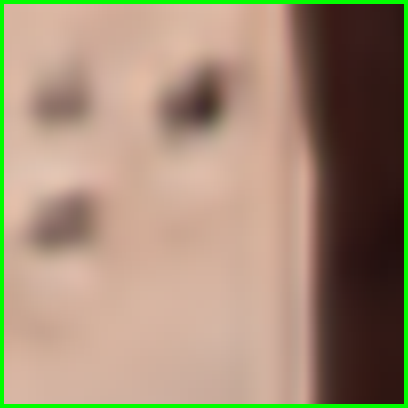}
\end{subfigure}
&
\begin{subfigure}{\cellasize}
\includegraphics[width=\imgsize]{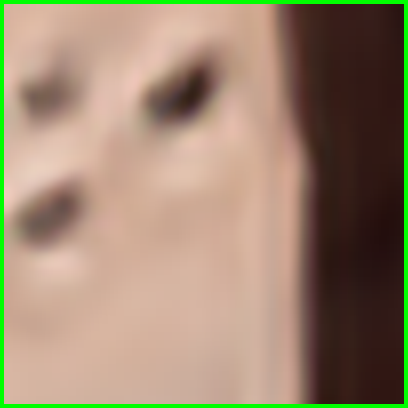}
\end{subfigure}
&
\begin{subfigure}{\cellasize}
\includegraphics[width=\imgsize]{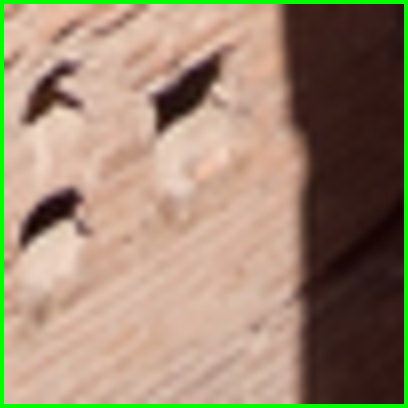}
\end{subfigure}
&
\begin{subfigure}{\cellasize}
\includegraphics[width=\imgsize]{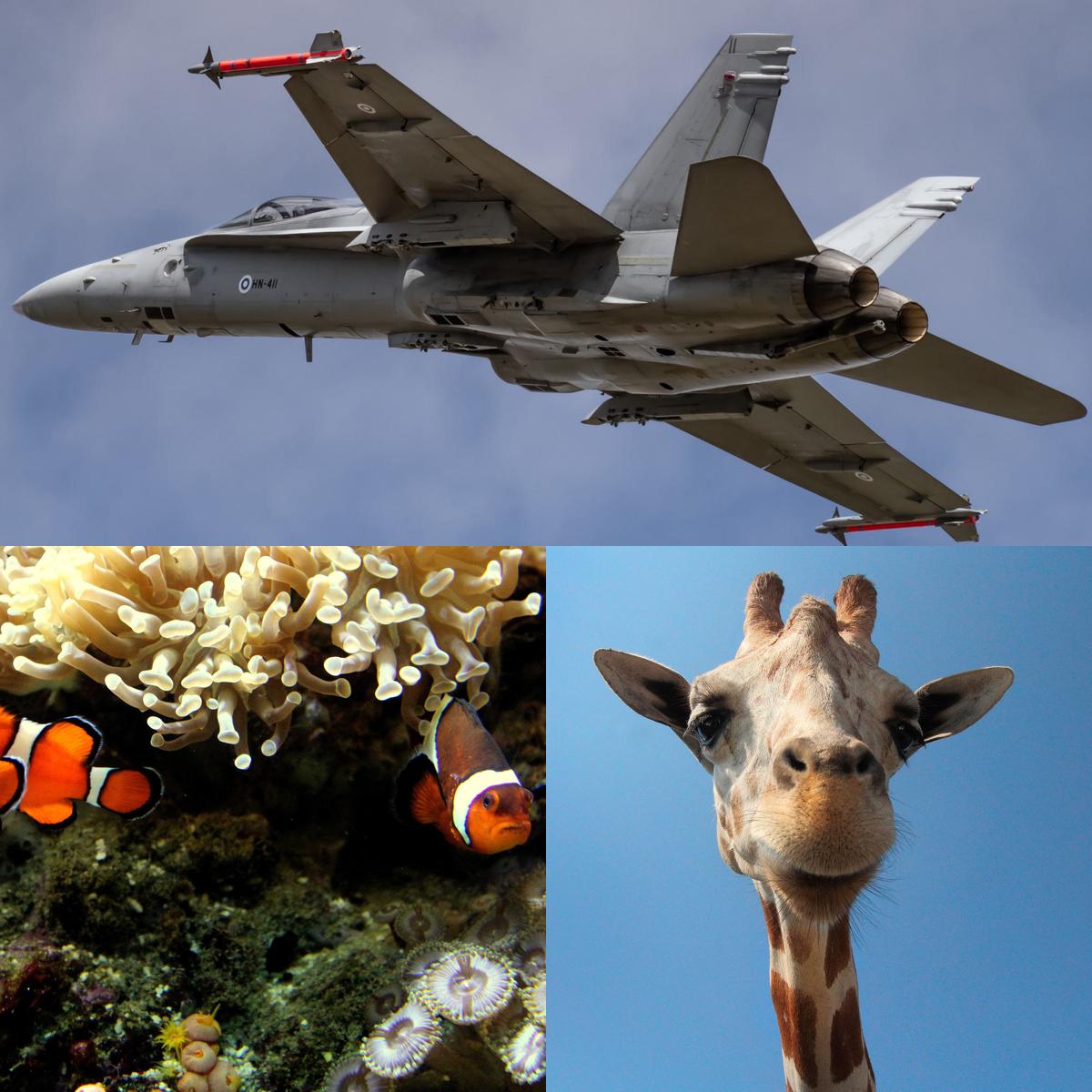}
\end{subfigure}
\\[\rowspace]

\end{tabular}
}
\endgroup
\caption{Qualitative comparison of DualSR with the reference add-on module. The first two rows display results for a scale of $\times$2, while the subsequent two rows show results for a scale of $\times$4.}
\label{fig:vision-ZSSR-based}
\vspace{-5pt}
\end{figure}

\subsubsection{Comparison of Reference Image Selection Methods.}
As mentioned earlier, the reference images employed in our methods may not inherently contain content relevant to the target images. Although the specific content of the reference images has minimal impact on our method, the selection of a reference image introduces subtle variations in the outcome of the target image. In response to this, we have devised an automatic selection method to choose the reference image from the high-resolution collection. This method involves calculating the mean values of the reference images in the RGB channels, respectively, and selecting the closest match to the target image. In Table \ref{tab:ref-selection}, we compare three distinct selection policies: auto-selection, random selection, and reverse auto-selection, wherein the farthest images are chosen as the reference images—an inverse approach to auto-selection. The results indicate that selecting HR images with similar color distribution (automatic selection method) leads to better SR outcomes.

\subsubsection{Comparison of the Number of Reference Images.}
Our method acquires domain knowledge from the provided reference HR images. To investigate the optimal number of reference images (denoted as $N$) required to impart sufficient domain knowledge, we conducted experiments using 1, 3, and 5 reference images. As depicted in Table \ref{tab:ref-count}, we observed a consistent trend where increasing the number of reference images correlates positively with improved SR performance. By increasing the number of reference images, the model gains deeper insights into a broader range of image characteristics, thereby leading to enhanced super-resolution (SR) results.

\subsection{Study of Reference-based Module on ZSSR Method}

With our proposed procedure for integrating HR images into the model, we extend this approach to zero-shot super-resolution (ZSSR) methods.
In our experiments, we integrate the reference add-on module into the DualSR \cite{DualSR} training network. The DIV2K dataset is used as our target image dataset, and we select reference images from DIV2KRK. The PSNR and SSIM values are reported in Table \ref{tab:exp-ZSSR-based}, which demonstrates the benefits of applying the reference add-on module. Fig. \ref{fig:vision-ZSSR-based} provides a visual comparison between our results and those obtained from DualSR, highlighting the effective reduction of artifacts achieved by our methods.

\begin{table}[!htbp]
    \caption{Quantitative analysis of the impact of the number (N) of reference images.}
    \label{tab:ref-count}
    \centering
    \scalebox{0.8}{
    \begin{tabular}{m{0.35\linewidth} m{0.1\linewidth} m{0.15\linewidth} m{0.15\linewidth} } \toprule
    Method \& Datasets & N & PSNR-Y & SSIM \\ \midrule
    \multirow{3}{*}{\makecell{DualSR + Ref. add-on\\DIV2KRK x2}} & 1  &  30.594 & 0.8551 \\
    & 3  &  30.880 & 0.8622 \\ 
    & 5  &  \textbf{30.918} & \textbf{0.8636} \\\bottomrule
    \end{tabular}
    }
\end{table}

\begin{table}
    \caption{Quantitative comparison of reference add-on module based on DualSR \cite{DualSR} on Track 2 of DIV2K.}
    \label{tab:exp-ZSSR-based}
    \centering
    \scalebox{0.8}{
    \begin{tabular}{m{0.4\linewidth} m{0.10\linewidth} m{0.15\linewidth} m{0.15\linewidth} } \toprule
    Method & Scale & PSNR-Y & SSIM \\ \midrule
    DualSR & $\times$2 & 26.011 & 0.7310  \\
    DualSR + Ref. add-on module & $\times$2 & \textbf{26.228} & \textbf{0.7353}  \\\midrule
    DualSR & $\times$4 & 23.079 & \textbf{0.6466}  \\
    DualSR + Ref. add-on module & $\times$4 & \textbf{23.230} & 0.6388  \\\bottomrule
    \end{tabular}
    }
\end{table}

\section{Conclusion}\label{sec:conclusion}
We present a novel and versatile approach to significantly enhance super-resolution (SR) performance by leveraging additional high-resolution images. Our method utilizes these supplementary images to facilitate learning a scale-specific degradation process. This approach distinguishes itself from conventional reference-based SR methods, which typically rely on reference images closely resembling the target image. Moreover, our proposed method is applicable in both well-trained blind SR and zero-shot SR scenarios. Additionally, our procedure offers flexibility in selecting the network backbone, enabling the substitution of kernel estimators and SR networks with more advanced techniques tailored specifically to the target image. Experimental results demonstrate that our specific instance, RDSR, surpasses baseline blind super-resolution algorithms in terms of both PSNR and SSIM, consistently delivering visually superior results across various scenarios.

\subsubsection{Acknowledgments}
This work was financially supported in part (project number: 112UA10019) by the Co-creation Platform of the Industry Academia Innovation School, NYCU, under the framework of the National Key Fields Industry-University Cooperation and Skilled Personnel Training Act, from the Ministry of Education (MOE) and industry partners in Taiwan.  It also supported in part by the National Science and Technology Council, Taiwan, under Grant NSTC-112-2221-E-A49-089-MY3, Grant NSTC-110-2221-E-A49-066-MY3, Grant NSTC-111- 2634-F-A49-010, Grant NSTC-112-2425-H-A49-001, and in part by the Higher Education Sprout Project of the National Yang Ming Chiao Tung University and the Ministry of Education (MOE), Taiwan. And we would like to express our gratitude for the support from MediaTek Inc.

%
%

\bibliographystyle{splncs04}
\bibliography{9-refs}



\clearpage
\section*{Supplementary}

Firstly, we introduce more details on the training process of our RDSR method, including hyperparammeter settings such as learning rate and the weights of the loss functions. Next, we explore the concept of designing a downsampling network. After that, we discuss further details about our reference add-on module based on ZSSR methods and our auto-selection reference image method. Finally, we present more visual results in Fig. \ref{fig:vision-BSR-supple} and Fig. \ref{fig:vision-ZSSR-supple}.

\section{Training Details of RDSR}
As mentioned above, the initial learning rate of downsampler $G_{dn}$ and upsampler $G_{up}$ are $2\times10^{-3}$ and $1\times10^{-5}$, respectively. Additionally, the initial learning rates of the degradation estimator $E_{k\ L{\rightarrow}H}$ and the discriminator $D_{up}$ is $5\times10^{-5}$ and $1\times10^{-5}$, respectively. Since our downsampling network is trained from scratch, we set a higher learning rate for it, and the upsampling network $G_{up}$ which includes both the degradation estimator $E_{k\ L{\rightarrow}H}$ and upscaler, is set for a lower initial rate.

Refer to previous research \cite{TTSR, C2Matching, MASA}, 
loss functions such as reconstruction loss, perceptual loss, and adversarial loss are commonly used in super-resolution task.
Here we applied $\mathcal{L}_{Charb}$ and $\mathcal{L}_{vgg}$
and set the same value 1 for Charbonnier loss \cite{LPSR} and perceptual loss \cite{perceptual2016}, taking into account the similarity of the pixel values and the perceptual similarity. Furthermore, we adopted the adversarial loss $\mathcal{L}_{gan}$ for the discriminator $D_{up}$ as mentioned in section 3.3. Both $\mathcal{L}_{gan}$ and $\mathcal{L}_{vgg}$ encourage our networks to produce more visually pleasing results.

\section{Criteria for Final Result}
During the Fine-tune Phase of our training, we conduct the evaluation process every 50 iterations to determine the optimal image for our training purposes. Specifically, we compute the cycle consistency loss for the entire input image. Additionally, we calculate BRISQUE\cite{brisque}, a no-reference quality assessment, on both our SR output image  and the super-resolution (SR) image generated by a pre-trained model at Initial Phase. If the cycle consistency loss is minimized and the score obtained from our SR output image surpasses that of the SR image, we preserve the image as our output result. This mechanism acts as a precautionary measure against the inclusion of corrupted images from our SR network. Conversely, if we fail to identify a superior image during the evaluation stage, we retain the original image as the output result. In our experiments, approximately 6.5\% of output images remain unchanged by our method, indicating that certain cases cannot be enhanced by our techniques.




\newcommand\celldsize{0.25\textwidth}
\newcommand\cellesize{0.23\textwidth}

\begin{figure}[!hb]
\small
\begingroup
\setlength{\tabcolsep}{0.2pt} 
\renewcommand{\arraystretch}{1} 

\begin{tabular}{ c c c c }

LR \centering & GT \centering & Ref. & ~
\\[\rowspace] 

\begin{subfigure}{\celldsize}
\includegraphics[width=\imgsize]{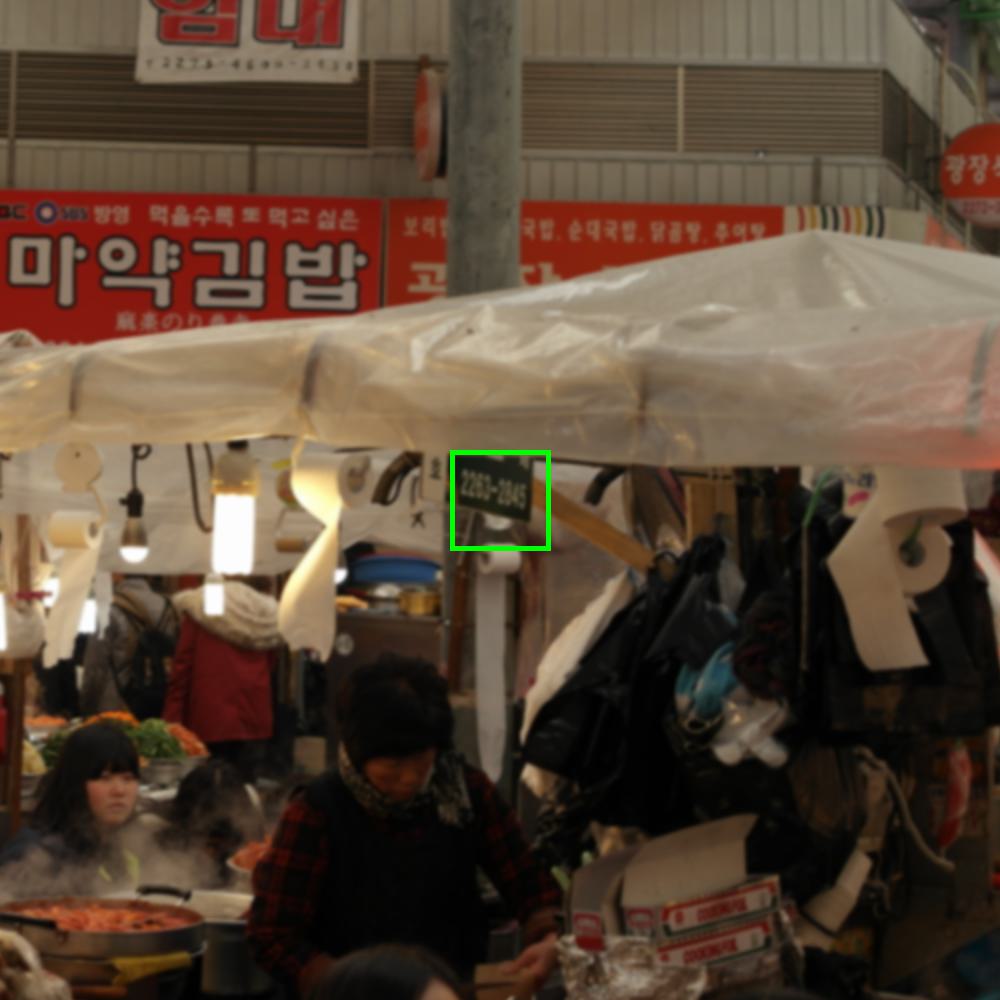}
\end{subfigure} 
&
\begin{subfigure}{\celldsize}
\includegraphics[width=\imgsize]{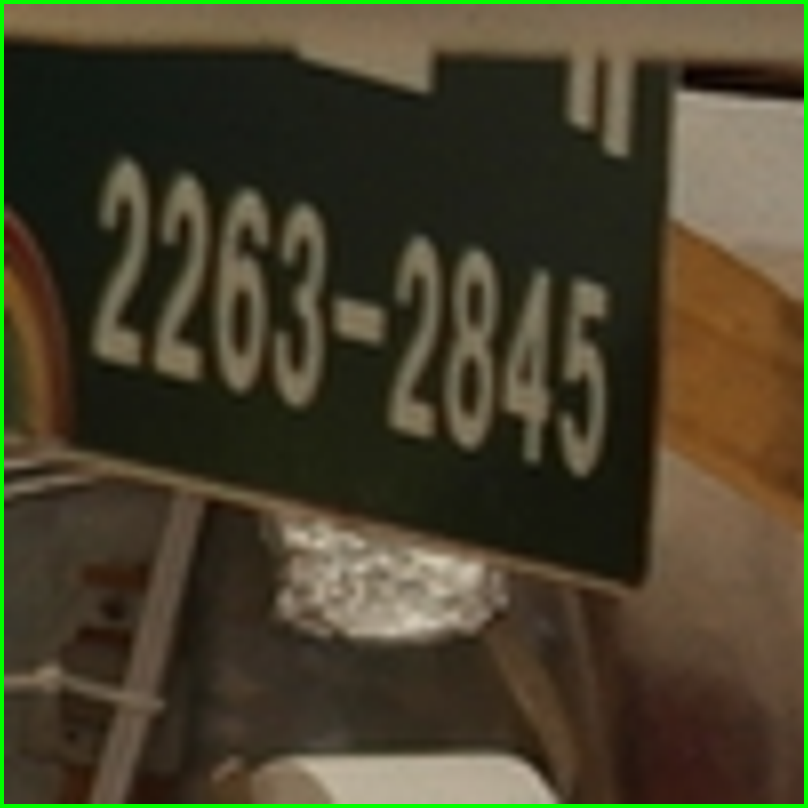}
\end{subfigure} 
&
\begin{subfigure}{\celldsize}
\includegraphics[width=\imgsize]{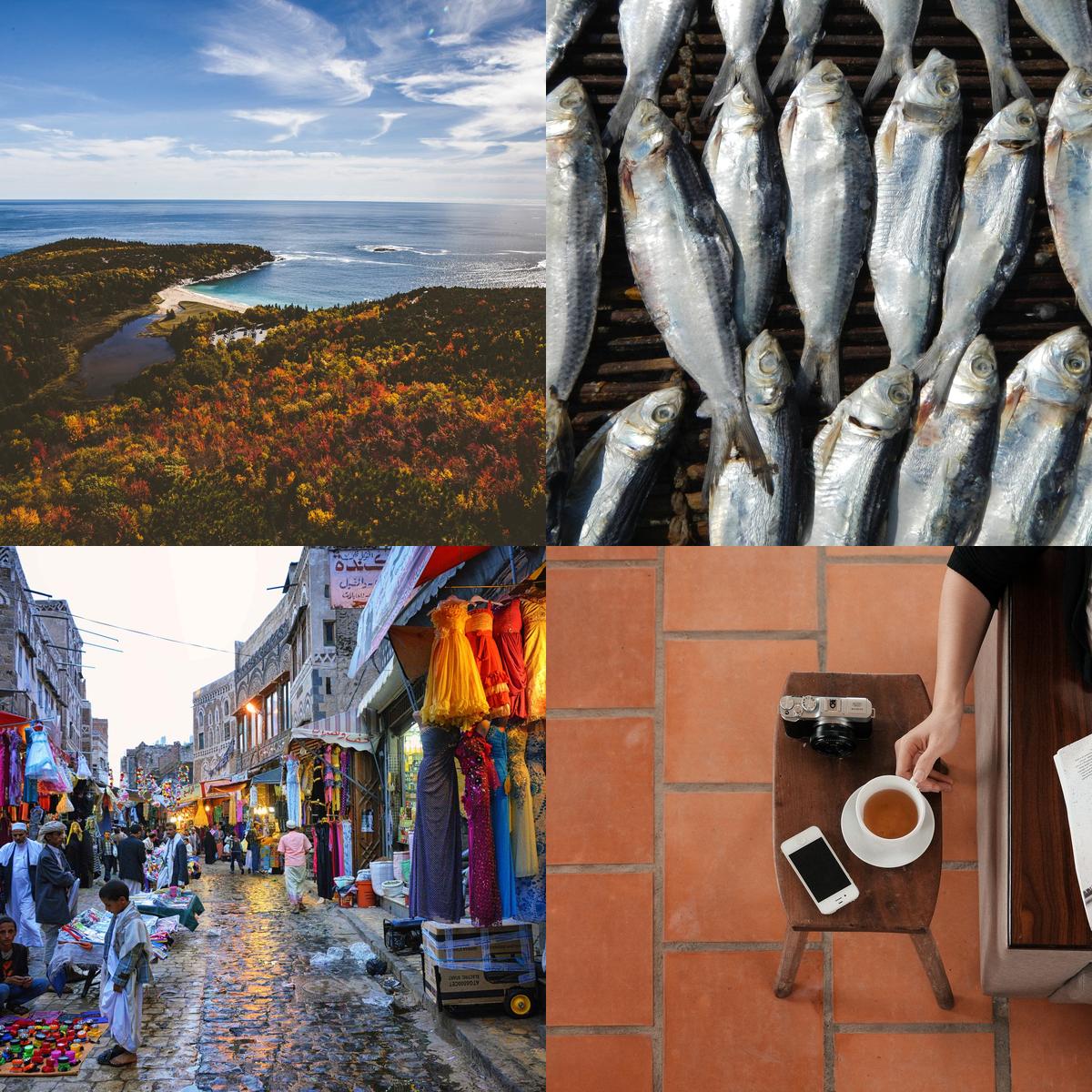}
\end{subfigure} 
&
~

\\[\rowspace]

Bicubic \centering & DASR \centering & DualSR \centering & RDSR(ours)
\\[\rowspace]
\begin{subfigure}{\celldsize}
\includegraphics[width=\imgsize]{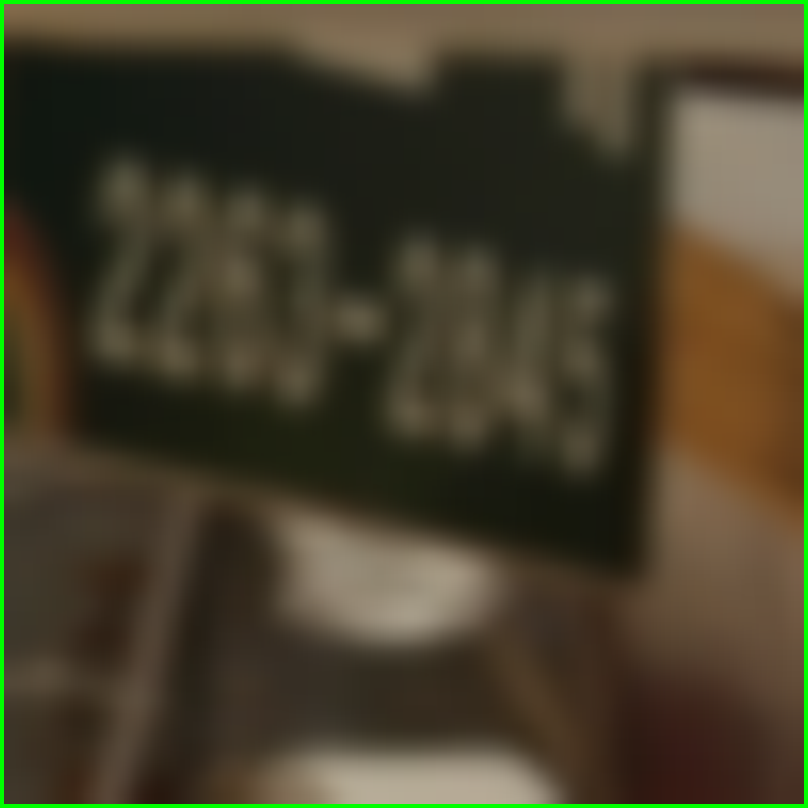}
\end{subfigure}
&
\begin{subfigure}{\celldsize}
\includegraphics[width=\imgsize]{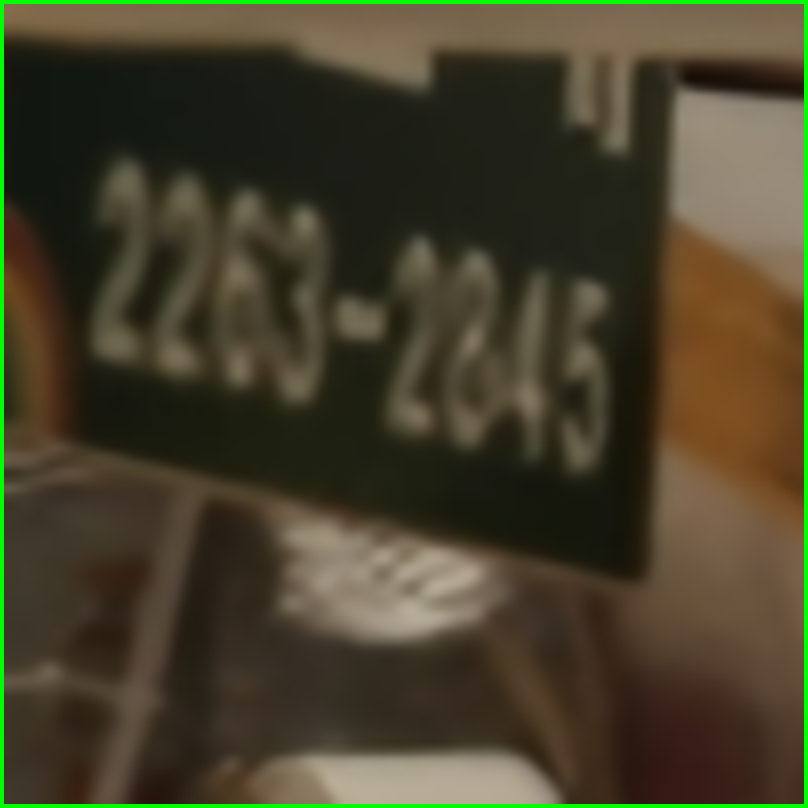}
\end{subfigure} 
&
\begin{subfigure}{\celldsize}
\includegraphics[width=\imgsize]{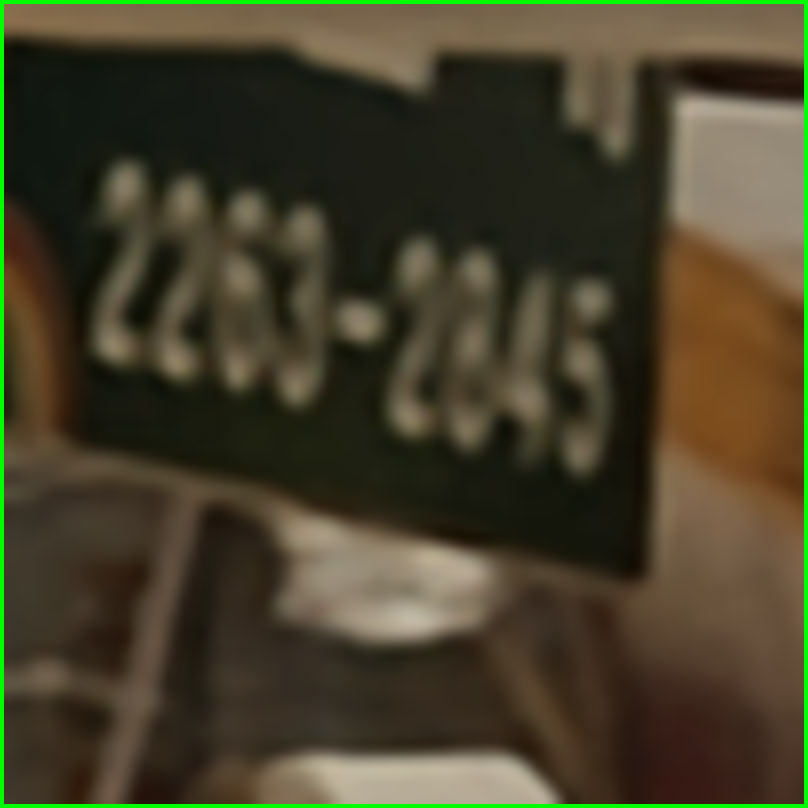}
\end{subfigure} 
&
\begin{subfigure}{\celldsize}
\includegraphics[width=\imgsize]{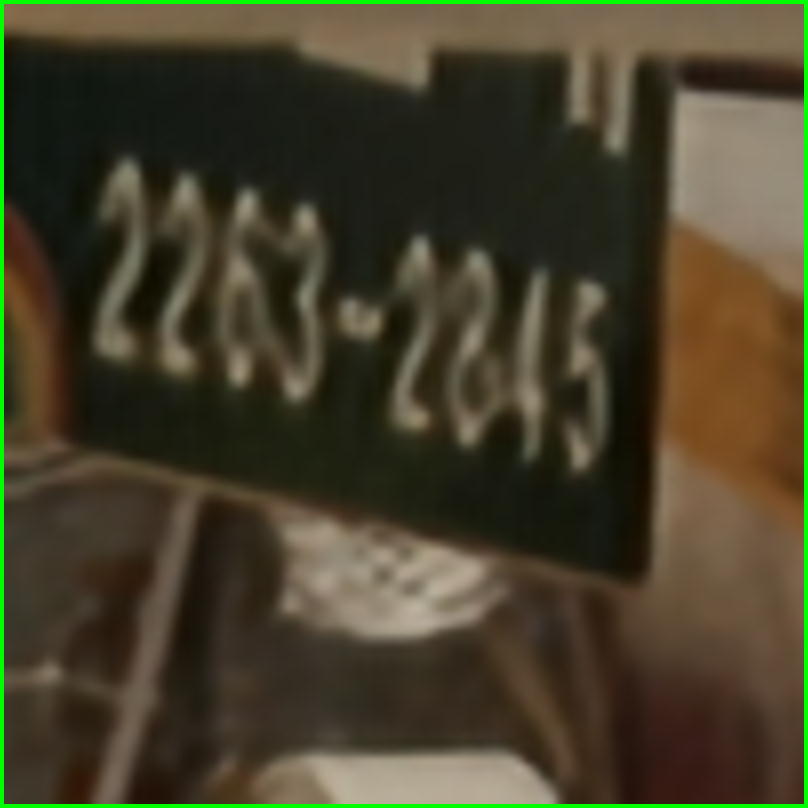}
\end{subfigure}

\\[\rowspace] 
\\[\rowspace] 
\\[\rowspace] 
LR \centering & GT \centering & Ref.  & ~ 
\\[\rowspace] 


\begin{subfigure}{\celldsize}
\includegraphics[width=\imgsize]{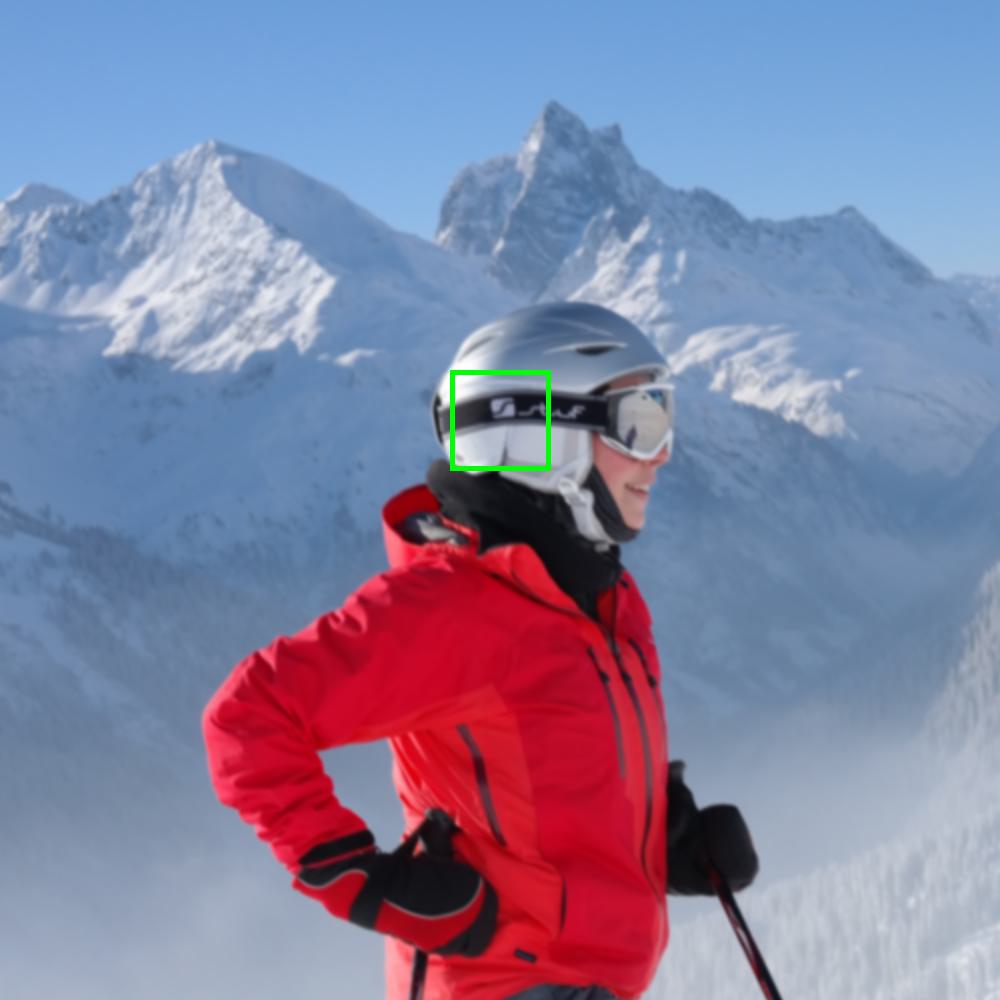}
\end{subfigure} 
&
\begin{subfigure}{\celldsize}
\includegraphics[width=\imgsize]{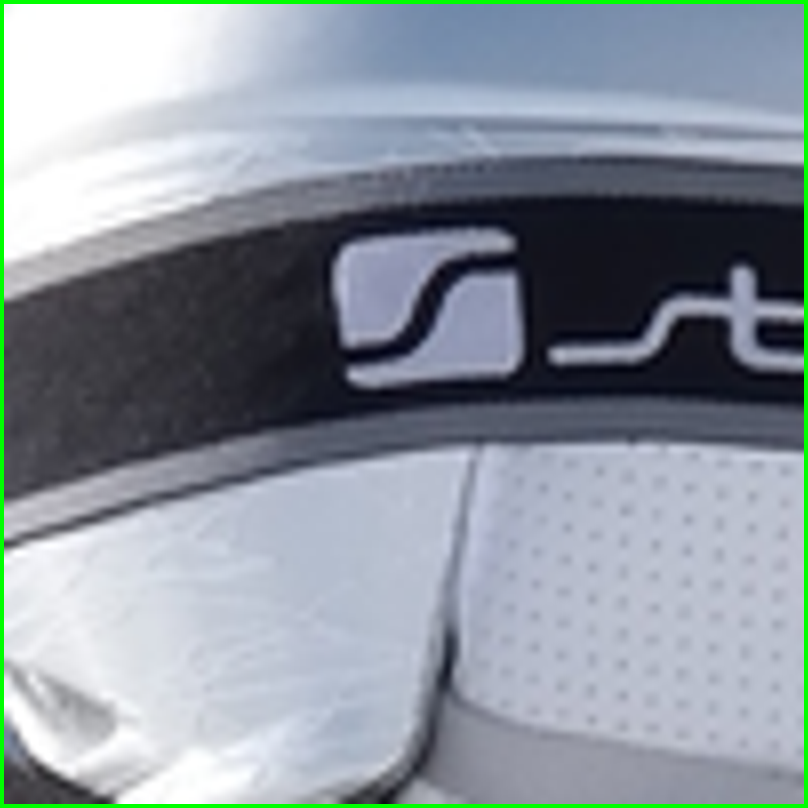}
\end{subfigure} 
&
\begin{subfigure}{\celldsize}
\includegraphics[width=\imgsize]{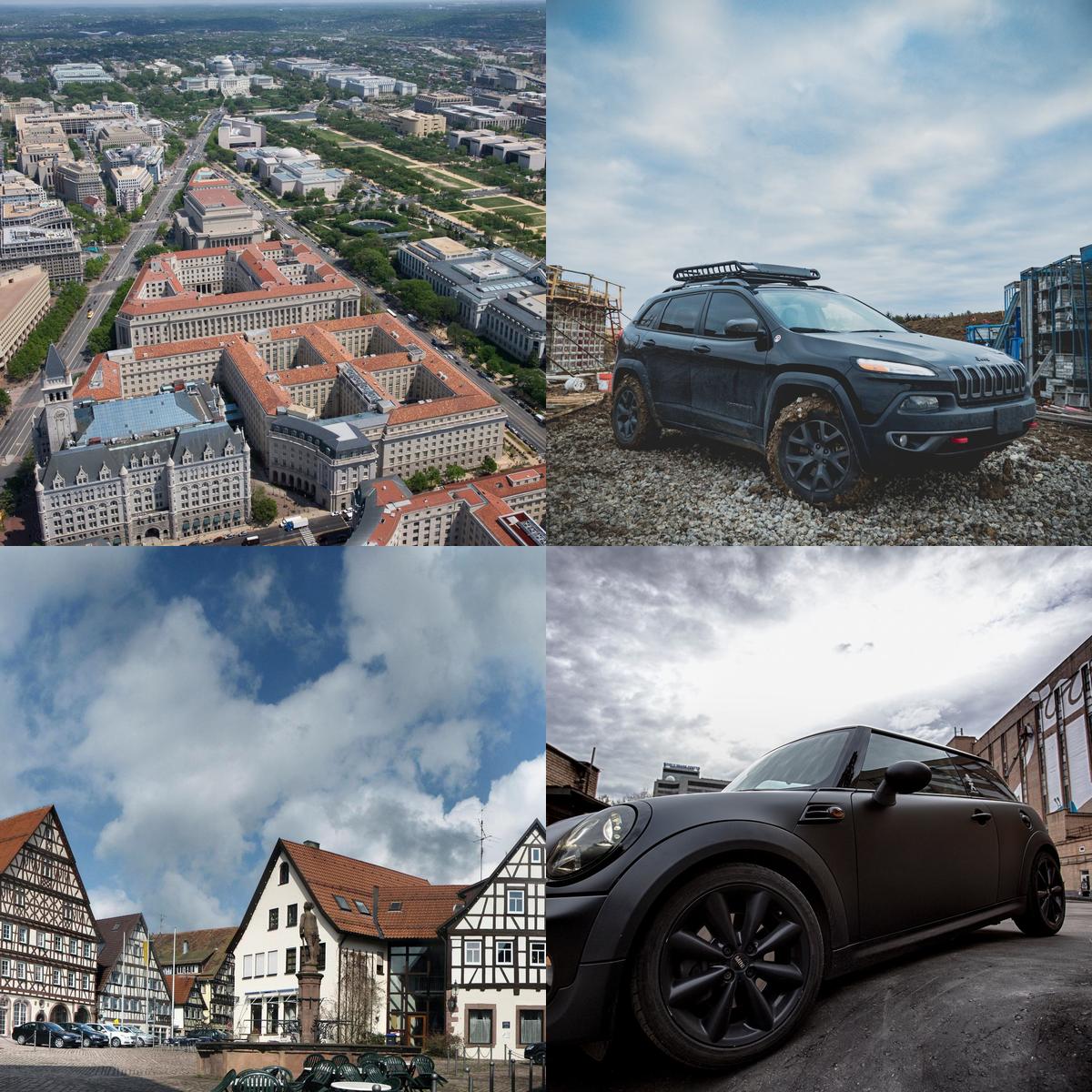}
\end{subfigure} 
&
~
\\[\rowspace] 

Bicubic \centering & DASR \centering & DualSR \centering & RDSR(ours)
\\[\rowspace]

\begin{subfigure}{\celldsize}
\includegraphics[width=\imgsize]{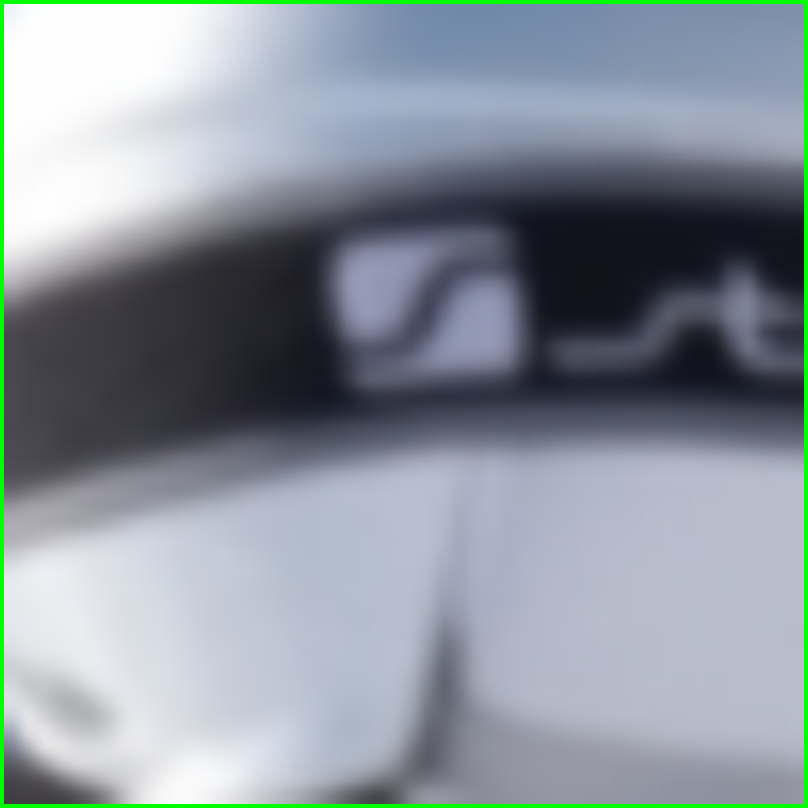}
\end{subfigure}
&
\begin{subfigure}{\celldsize}
\includegraphics[width=\imgsize]{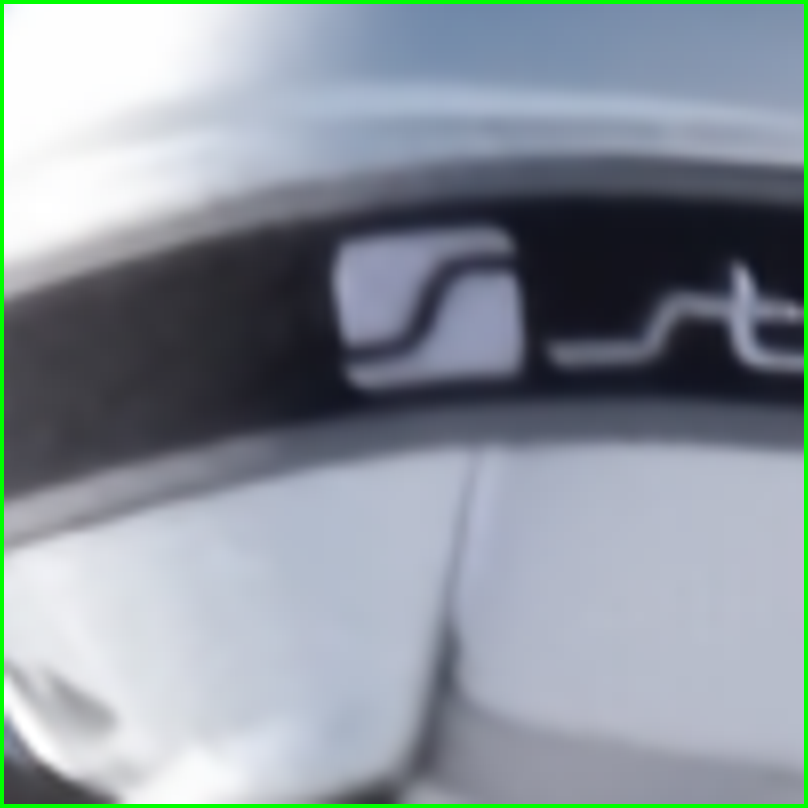}
\end{subfigure} 
&
\begin{subfigure}{\celldsize}
\includegraphics[width=\imgsize]{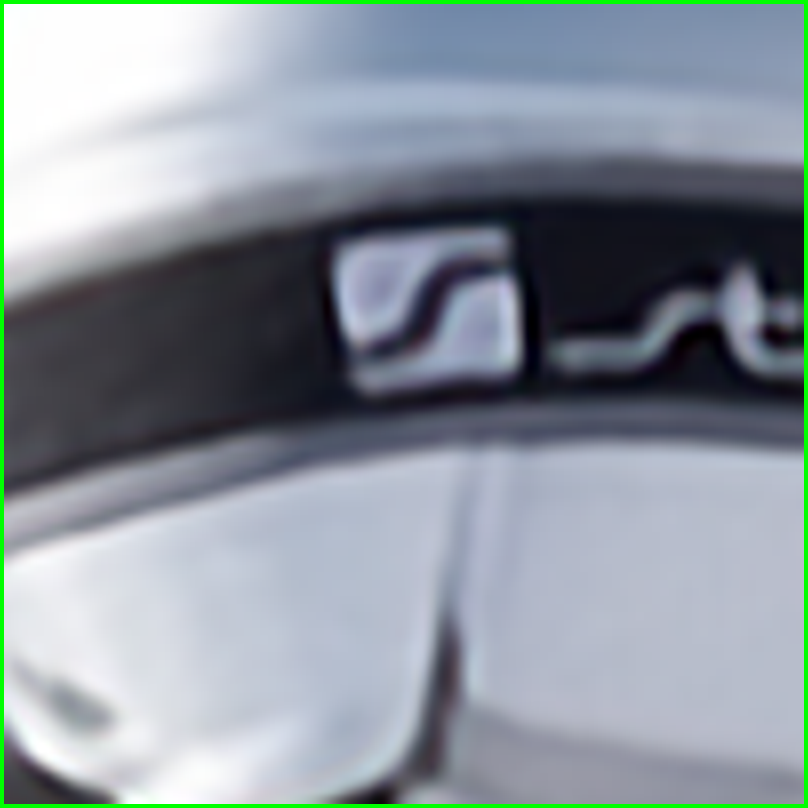}
\end{subfigure} 
&
\begin{subfigure}{\celldsize}
\includegraphics[width=\imgsize]{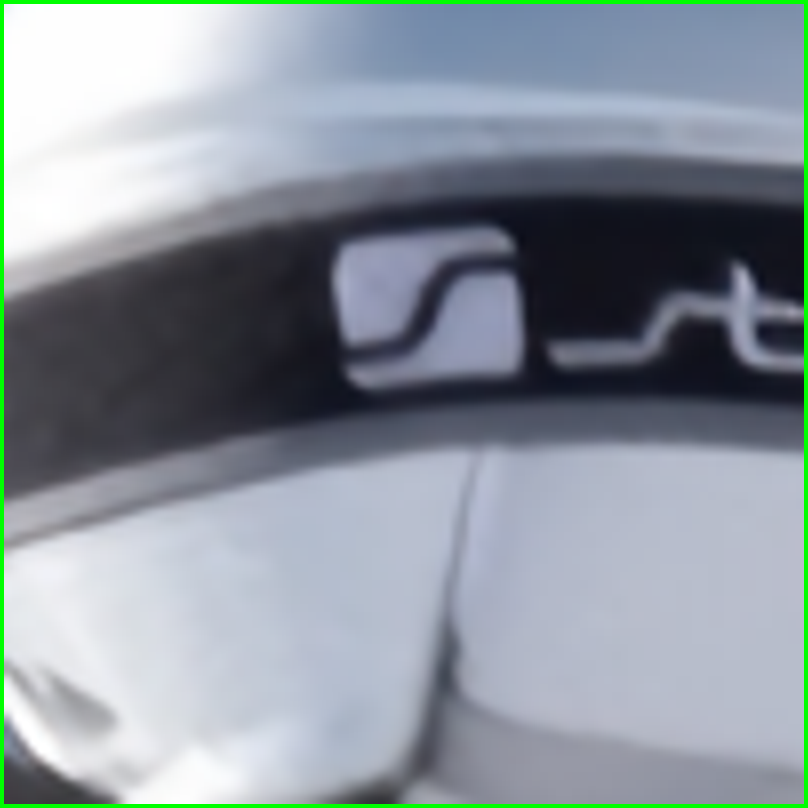}
\end{subfigure}

\\[\rowspace] 

\end{tabular}
\endgroup
\caption{Qualitative comparison of the RDSR based on "Blind Super Resolution" method with the isotropic kernel at scale $\times$2.}
\label{fig:vision-BSR-supple}
\end{figure}

\newcommand\cellbsize{0.33\textwidth}
\newcommand\cellcsize{0.30\textwidth}

\begin{figure}[!hb]
\small
\begingroup
\setlength{\tabcolsep}{0.2pt} 
\renewcommand{\arraystretch}{1} 


\begin{tabular}{ c c c}
LR \centering & GT \centering & Ref \\
[\rowspace] 

\begin{subfigure}{\cellcsize}
\includegraphics[width=\imgsize]{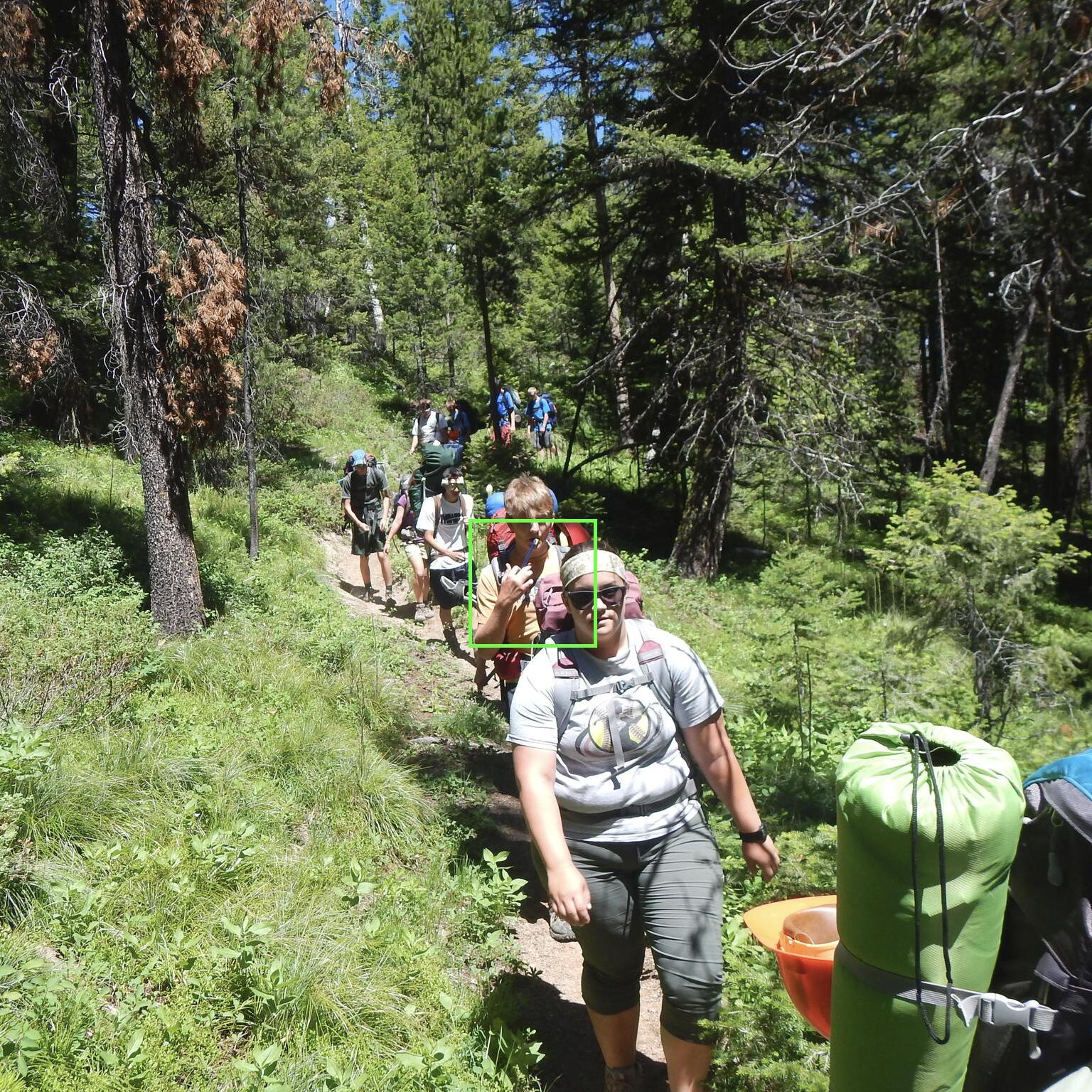}
\end{subfigure} 
&
\begin{subfigure}{\cellcsize}
\includegraphics[width=\imgsize]{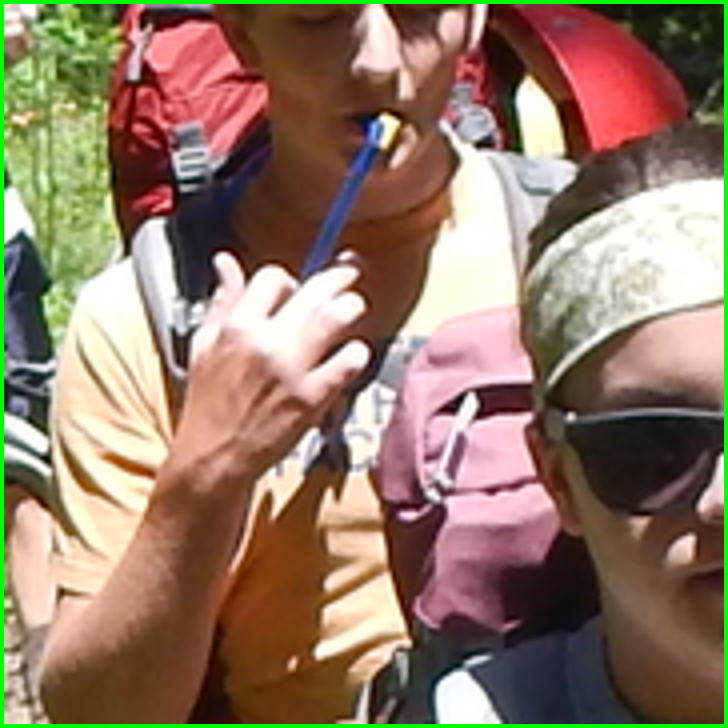}
\end{subfigure}
&
\begin{subfigure}{\cellcsize}
\includegraphics[width=\imgsize]{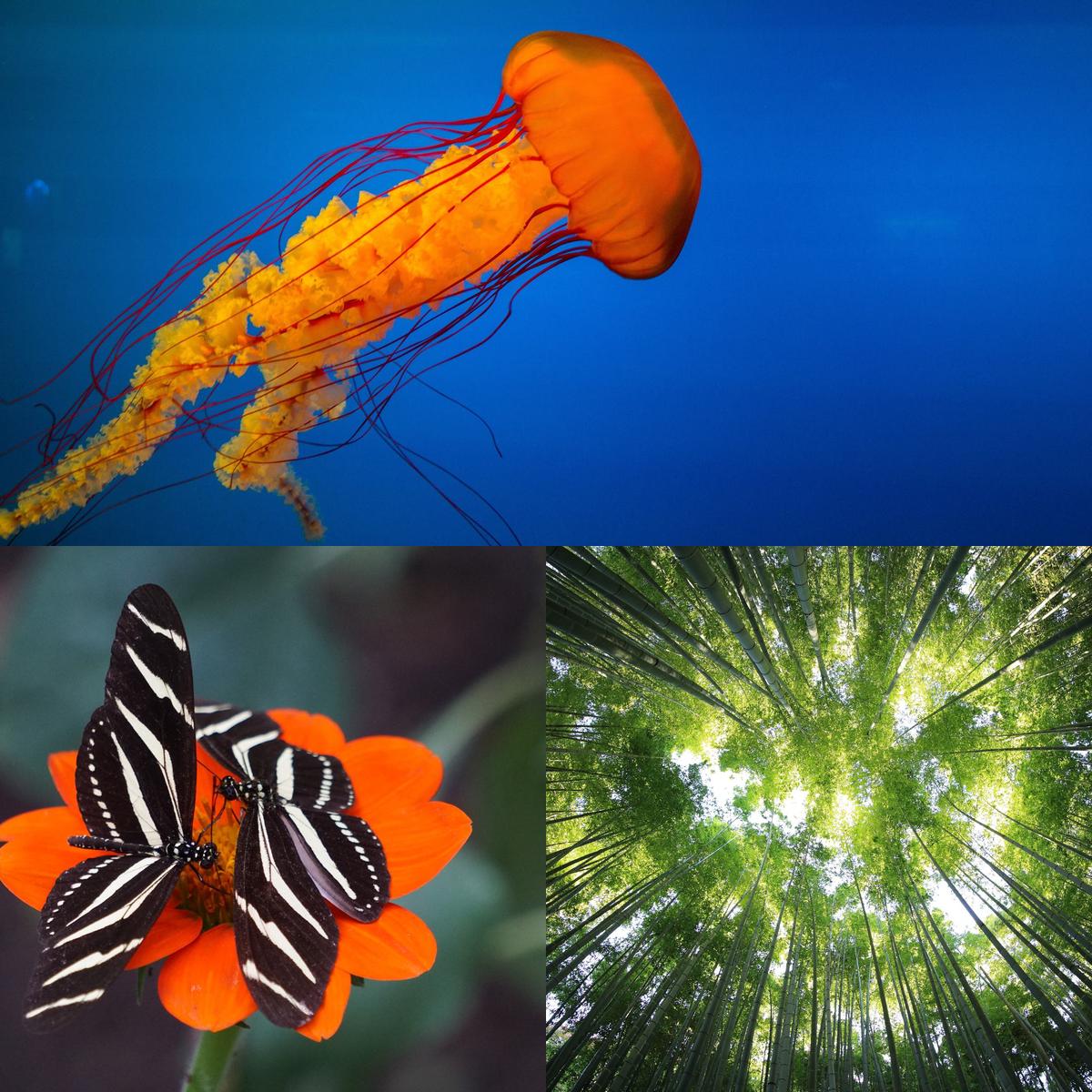}
\end{subfigure} 

\\[\rowspace] 
Bicubic \centering & DualSR \centering &  Ref add-on(ours)  \\

\begin{subfigure}{\cellcsize}
\includegraphics[width=\imgsize]{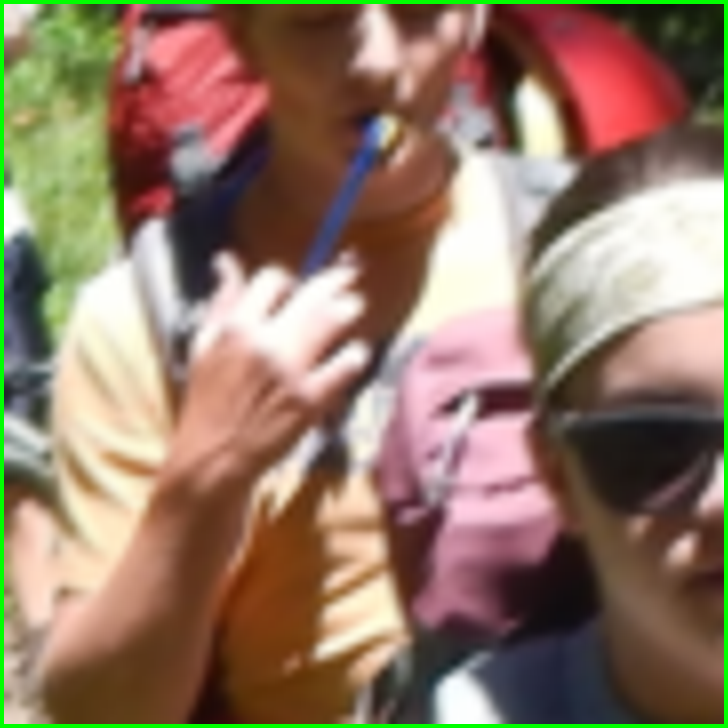}
\end{subfigure} 
&
\begin{subfigure}{\cellcsize}
\includegraphics[width=\imgsize]{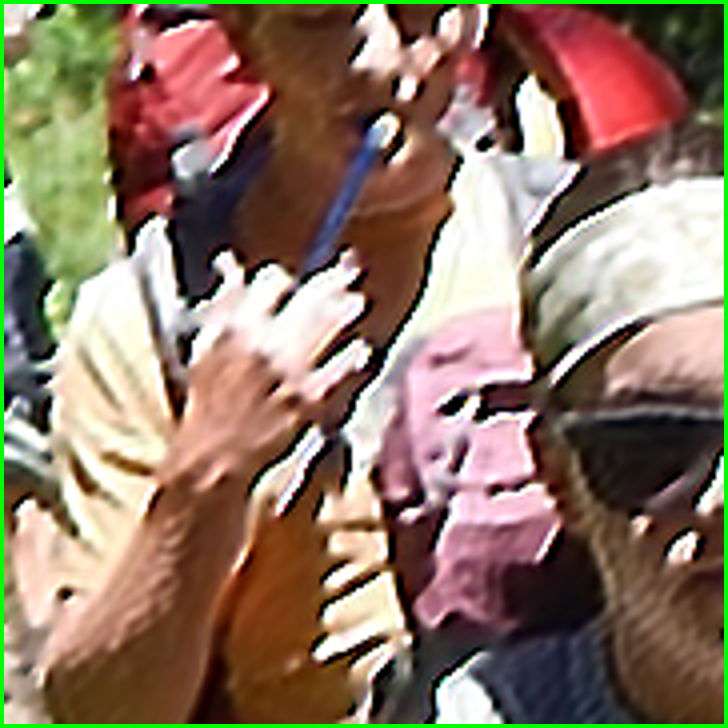}
\end{subfigure}
&
\begin{subfigure}{\cellcsize}
\includegraphics[width=\imgsize]{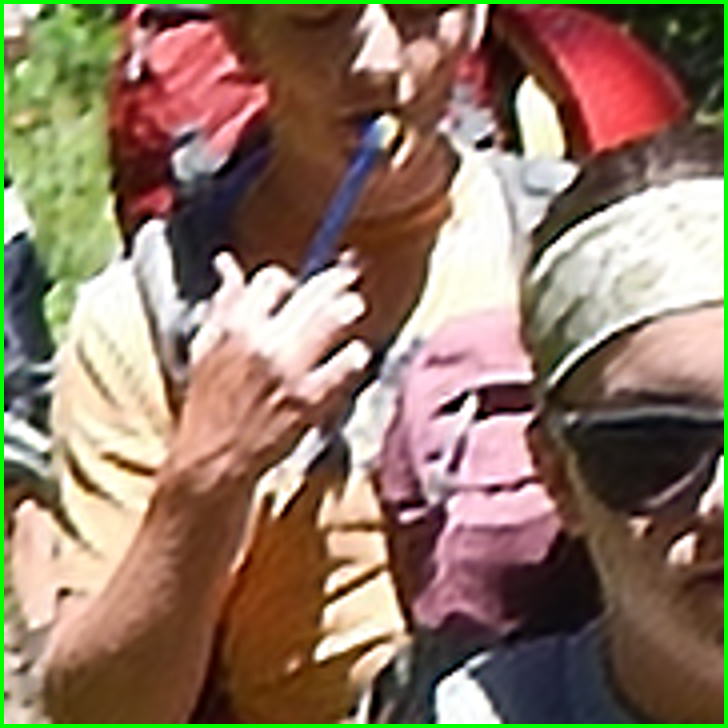}
\end{subfigure} 

\\ \midrule [\rowspace] 
\\[\rowspace] 
\\[\rowspace] 
LR \centering & GT\centering & Ref \\
[\rowspace]
\begin{subfigure}{\cellcsize}
\includegraphics[width=\imgsize]{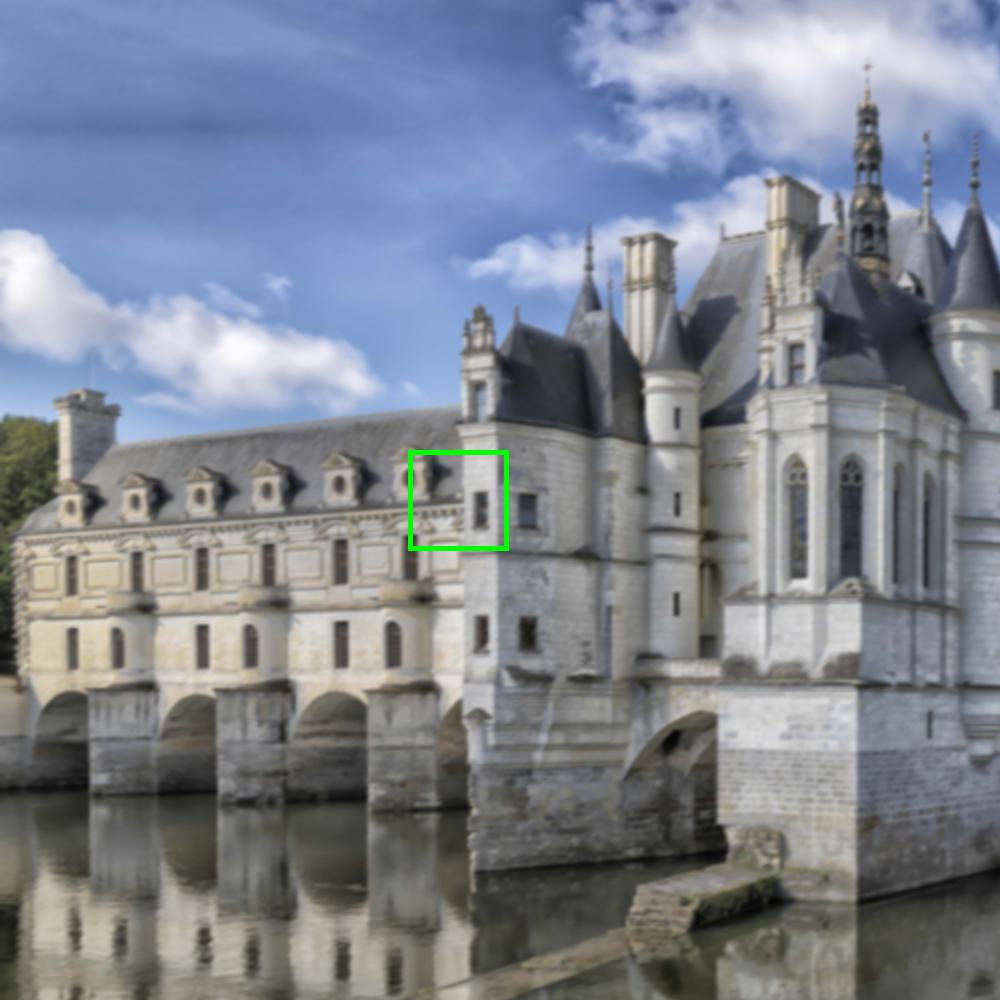}
\end{subfigure} 
&
\begin{subfigure}{\cellcsize}
\includegraphics[width=\imgsize]{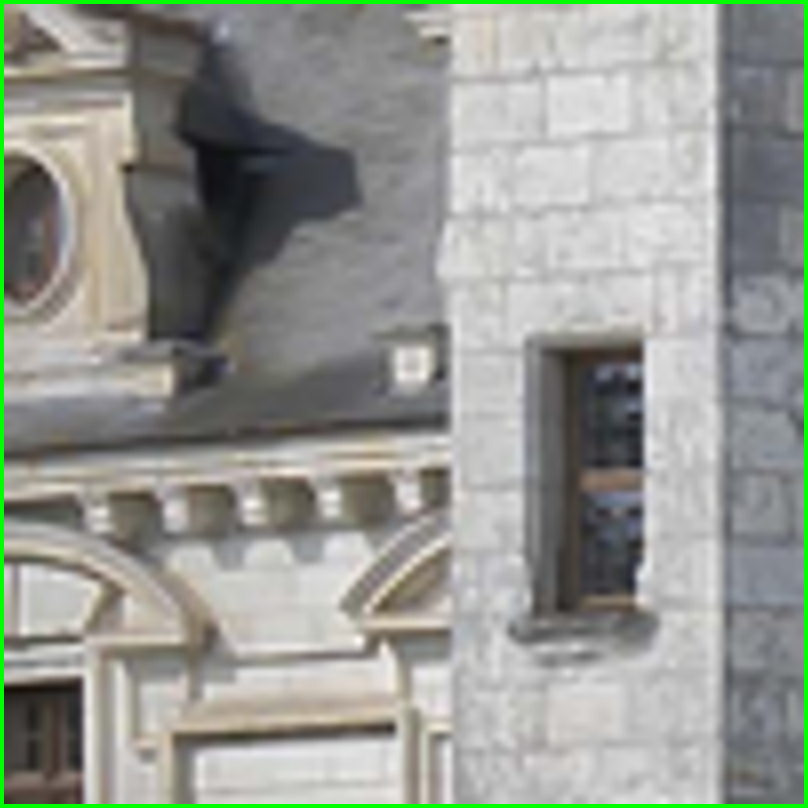}
\end{subfigure}
&
\begin{subfigure}{\cellcsize}
\includegraphics[width=\imgsize]{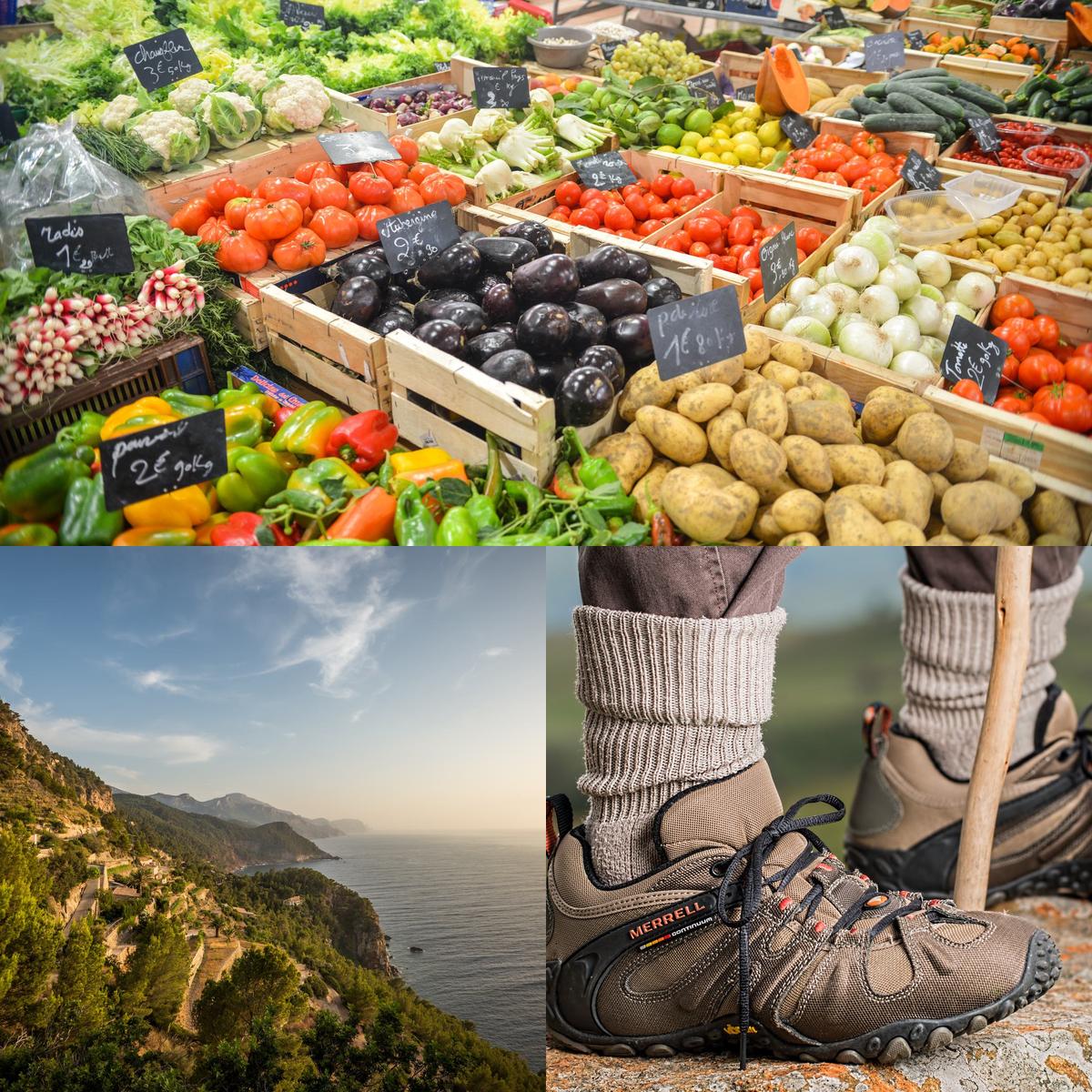}
\end{subfigure} 

\\[\rowspace]

Bicubic \centering & DualSR \centering &  Ref add-on(ours)  \\
\begin{subfigure}{\cellcsize}
\includegraphics[width=\imgsize]{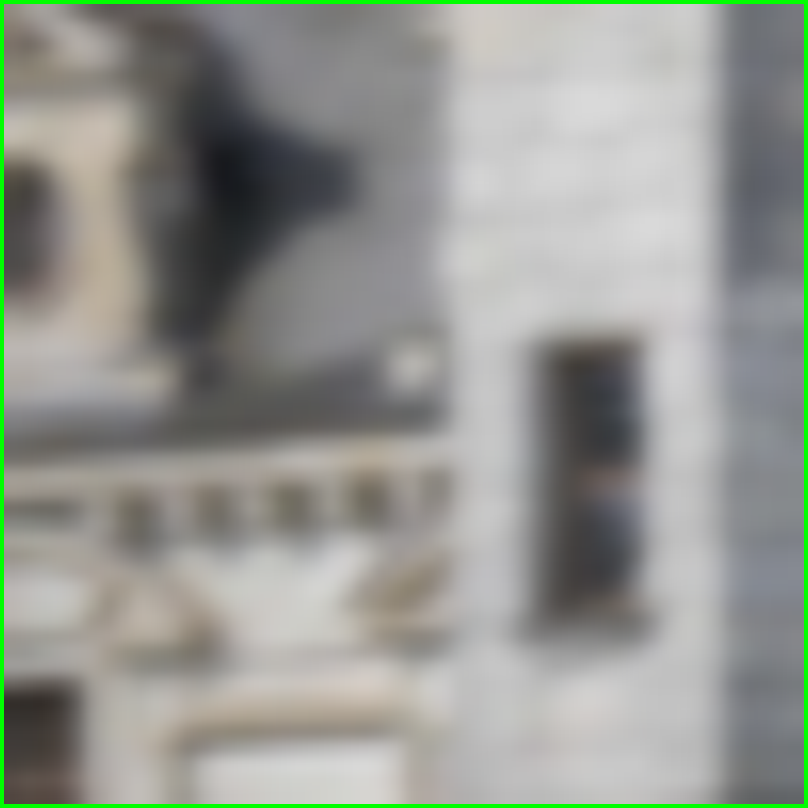}
\end{subfigure} 
&
\begin{subfigure}{\cellcsize}
\includegraphics[width=\imgsize]{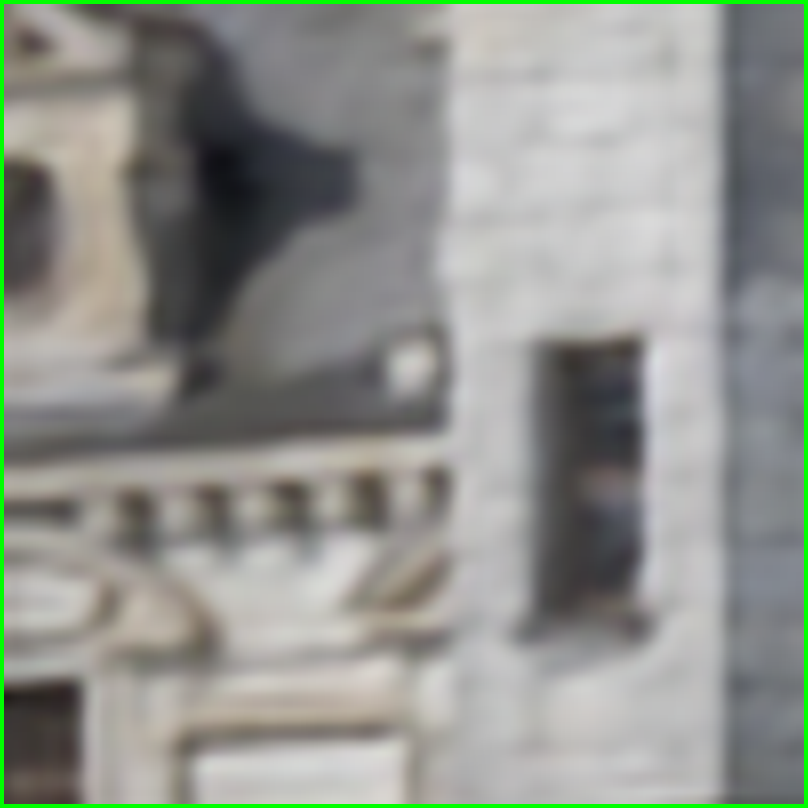}
\end{subfigure}
&
\begin{subfigure}{\cellcsize}
\includegraphics[width=\imgsize]{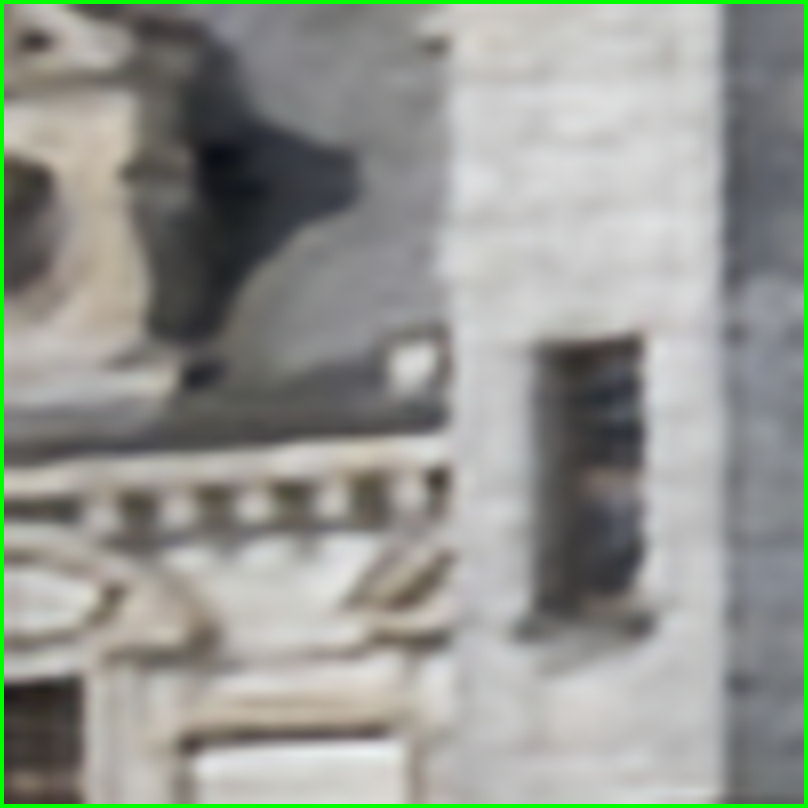}
\end{subfigure}



\\[\rowspace] 

\end{tabular}
\endgroup
\caption{Qualitative comparison of DualSR with the reference add-on module at scale $\times$2.}
\label{fig:vision-ZSSR-supple}
\end{figure}

\section{Design of Downsampling Network}
We adopt the methodology of KernelGAN \cite{KernelGAN} to formulate our downsampling network. This framework enables the extraction of an explicit kernel from the weights of deep linear networks, allowing for the direct computation of the kernel prior from downsampling networks $G_{dn}$. For instance, we utilize the architecture of $G_{dn}$ resembling KernelGAN \cite{KernelGAN}, comprising 5 linear convolutional layers with the first 3 filter are 7, 5, and 3, and the rest are 1. Consequently, this setup yields an explicit kernel with a receptive field size of 13, indicating the network's ability to accommodate linear degradation processes within a kernel size of 13. Hence, if the kernel size of the input low-resolution (LR) image exceeds $13\times13$, it is advisable to adjust the architecture of $G_{dn}$.

The downsampling network can be trained using either supervised or unsupervised methods. In the supervised method, akin to Section 3.2, $G_{dn}$ is trained directly with high-resolution (HR) images from a pre-trained model. Conversely, the unsupervised method, exemplified by KernelGAN \cite{KernelGAN} as utilized in the reference add-on module based on ZSSR, provides an alternative approach considering various degradation conditions or image characteristics. In our experimental in section 4.4, we have determined that the supervised approach outperforms the unsupervised one. Hence, we opt to train the downsampling network $G_{dn}$ directly, based on this rationale.

\section{Training details of Reference Add-on Module}

We will provide more details on how to integrate our reference add-on module into DualSR, as mentioned previously.  
Similar to RDSR method, our process includes Initial Phase and Fine-tune Phase.
Initially, we refrained from directly activating the reference add-on module and instead ran pure DualSR for a set number of iterations to obtain the preliminary degradation process specific to the input target image.
Enabling the reference branch from the start could potentially introduce a distinct degradation process compared to the input target image due to the intervention of the reference branch. In our implementation, the Fine-tune Phase begins at 1000 iterations and concludes at 2500 iterations, resulting in a total of 3000 training iterations.
Notably, through monitoring the DualSR training process, we observed that the degradation kernel stabilizes after 1000 training iterations, leading us to select 1000 iterations for the Initial Phase. Additionally, we provide supplementary visual results utilizing the reference add-on module based on DualSR, as depicted in Fig. \ref{fig:vision-ZSSR-supple}.

\section{Detail of Auto-select Reference Image Method}

The auto-selection method process is summarized in Algorithm \ref{alg:auto-selection}. We split the RGB channels from the images and calculate the mean values for the input target image. Then the mean for each reference image is to compute MSE with the input target images. Finally, we sort and take the top few reference images as our selection result. Using this approach, we have the ability to choose reference images from a pool of high-resolution images, surpassing the efficacy of random selection.


\renewcommand{\algorithmicrequire}{\textbf{Input:}}
\renewcommand{\algorithmicensure}{\textbf{Output:}}

\begin{algorithm}
    \caption{Reference images selection from a HR collection}
    \label{alg:auto-selection}
    \begin{algorithmic}[1]
        \Require
            {$X_{LR}$,  
            $Ref_{HR}^{*} = \{R_{1}, R_{2}, ..., R_{n}\}$}
        \Ensure
            {$Ref_{mse}^{*} = \{R_{1-mse}, R_{2-mse}, ..., R_{n-mse}\}$}
        \State
        {$Ref_{mse}^{*} = \{\}$}
        \State 
        {$X_{r}$, $X_{g}$, $X_{b}$ = \texttt{split\_channel}($X_{LR}$)}
        \State 
        {$\overline{X_{r}}$, $\overline{X_{g}}$, $\overline{X_{b}}$ = \texttt{mean}($X_{r}$, $X_{g}$, $X_{b}$)}
        \For{$i$ $\gets$ $1$ to $n$}
        \State 
        {$R_{i-r}$, $R_{i-g}$, $R_{i-b}$ = \texttt{split\_channel}($R_{i}$)}
        \State
        {$\overline{R_{i-r}}$, $\overline{R_{i-g}}$, $\overline{R_{i-b}}$ = \texttt{mean}($R_{i-r}$, $R_{i-g}$, $R_{i-b}$)}
        \State
        {$R_{i-mse}$ = $(\overline{X_{r}} - \overline{R_{i-r}})^{2}$ + $(\overline{X_{g}} - \overline{R_{i-g}})^{2}$ + $(\overline{X_{b}} - \overline{R_{i-b}})^{2}$}
        \State {$Ref_{mse}^{*}$ $\gets$ $R_{i-mse}$}
        \EndFor
        \State
        \texttt{sort}($Ref_{mse}^{*}$)
        \State {return $Ref_{mse}^{*}$}
    \end{algorithmic}
\end{algorithm}


\end{document}